\documentclass[lettersize,journal]{IEEEtran}
\usepackage{amsmath,amsfonts}
\usepackage{algorithmic}
\usepackage{algorithm}
\usepackage{array}
\usepackage[caption=false,font=normalsize,labelfont=sf,textfont=sf]{subfig}
\usepackage{textcomp}
\usepackage{stfloats}
\usepackage{url}
\usepackage{verbatim}
\usepackage{graphicx}
\usepackage{cite}
\ifCLASSOPTIONcompsoc
  \usepackage[nocompress]{cite}
\else
  \usepackage{cite}
\fi

\ifCLASSINFOpdf
\else
\fi

\usepackage{amssymb}
\usepackage{amsmath}

\usepackage[table]{xcolor}
\usepackage{colortbl}
\usepackage{longtable}
\usepackage{multirow}
\usepackage{textcomp}
\usepackage{makecell}
\usepackage{graphicx}
\usepackage{hyperref}
\hypersetup{
    colorlinks=true,
    citecolor=blue,      
    linkcolor=red,       
    urlcolor=cyan,
    breaklinks=true
}

\usepackage{hyphenat}        
\definecolor{dimgray}{RGB}{230, 230, 230}
\definecolor{lightgray}{RGB}{245,245,245}
\definecolor{uafcolor}{RGB}{255,230,230}
\definecolor{oobcolor}{RGB}{230,255,230}
\definecolor{racecolor}{RGB}{255,255,200}
\definecolor{typecolor}{RGB}{230,240,255}
\definecolor{intcolor}{RGB}{255,240,230}

\definecolor{successgreen}{RGB}{200,255,200}
\definecolor{warningyellow}{RGB}{255,250,200}
\definecolor{failurered}{RGB}{255,220,220}

\newcommand{\cmark}{$\checkmark$}  
\newcommand{\xmark}{$\times$}      
\newcommand{\wmark}{$\sim$}        

\usepackage{tikz}
\usepackage{amsmath}

\usepackage{filecontents}

\begin{document}

\title{\Large \bf \textsc{PrimSynth}: An Agentic Approach to Discover, Validate, and Synthesize Exploit Primitives for Linux Kernel Vulnerabilities}

\author{Pengfei Wang\textsuperscript{1}, Anying Chen\textsuperscript{2}, Danjun Liu\textsuperscript{1}, Xu Zhou\textsuperscript{1}, Wei Xie\textsuperscript{1} \\ 
{\small1. National University of Defense Technology, 2. Clouditera}
\thanks{}
\thanks{}
}

\markboth{IEEE Transactions on Information Forensics and Security}%
{Pengfei Wang \MakeLowercase{\textit{et al.}}:\textsc{PrimSynth}: An Agentic Approach to Discover, Validate, and Synthesize Exploit Primitives for Linux Kernel Vulnerabilities}


\maketitle

\begin{abstract}
Linux kernel vulnerabilities are critical to downstream systems. Despite extensive research on automated kernel exploitation, a fundamental challenge remains the conceptual gap between abstract exploit strategies and concrete technical operations. To fill this gap, this paper introduces a systematic characterization that formalizes six classes of exploit primitives from logical capability to validatable effect. Then, an extended exploit strategy representation is proposed, which couples primitive upgrading strategies with primitive path code synthesis rules governing object constraints, temporal sequencing, environment prerequisites, and validation constraints. 
Building upon this foundation, this paper presents \textsc{PrimSynth}, a multi-agent framework that encapsulates these representations through coordinated agents to discover, validate, and synthesize exploit primitives for memory corruption vulnerabilities in the Linux kernel. These agents operate in an iterative closed loop until valid primitives are found, leveraging validation signals as evidence of exploitable state transitions to ground primitive synthesis decisions. An automated method for extracting and validating primitives is also proposed based on vulnerability-directed execution and a rebootable validation environment.
\textsc{PrimSynth} is evaluated on 16 real-world Linux kernel CVEs spanning 5 vulnerability types. Experimental results show that \textsc{PrimSynth} achieves reliable primitive extraction, maintaining a 100\% primitive match
rate. For primitive synthesis, \textsc{PrimSynth} successfully synthesizes multi-primitive exploitation chains with 82.4\% strategy synthesis rate (SSR) when the public PoC is available and a 61.3\% SSR without the guidance of
primitive hypotheses. For efficiency, \textsc{PrimSynth} completes end-to-end primitive synthesis in 72.4s on average, significantly outperforming semi-automated analysis or manual analysis that require hours to days. 
\end{abstract}


\section{Introduction}

The Linux kernel serves as the upstream source for all open-source or commercial distributions, and its security vulnerabilities propagate into downstream systems governing cloud infrastructure, web services, and embedded devices. Therefore, research on systematic reproduction and automated exploitability assessment of kernel vulnerabilities is proliferating \cite{dedup,hundreds,yome}.

Automated exploitation of the Linux kernel has been a long-standing open problem. The main challenge lies in the gap between the conceptual exploit strategies and the procedurally technical operations. Much of this diverse and nuanced knowledge resides only in the minds of human experts.
To bridge the knowledge gap in Linux kernel exploitation techniques, the concept of exploit primitives has been proposed. An \textit{exploit primitive} refers to a program state that violates security requirements and enables an attacker to gain capabilities beyond those provided by the original program. By providing a unified and formal framework, exploit primitives make vulnerability exploitation analysis more systematic and comparable.


Current research on exploit primitives typically targets a single vulnerability type or a single exploitation phase. As a result, operations used to realize one primitive are difficult to transfer into reproducible primitives in subsequent chainable phases. In addition, prior research on primitive mainly emphasizes target-object discovery and heap layout manipulation, but lacks a general validation methodology that can reliably localize candidate primitives and support further identification and extraction. Existing primitive analysis tends to focus on vulnerability-triggering paths, while the broader impact and composability of reproducible primitives are rarely evaluated because such assessments are labor-intensive. Although fast-growing LLM (Large Language Model) driven agents can generate PoCs for specific vulnerabilities \cite{Prompt2Own}, the process is consumptive, and the results are unstable without formal guidance on primitive composition.
Therefore, automated exploit generation (AEG) still faces challenges due to the lack of: (i) systematic characterization of validated and potential primitives; (ii) a general method to determine primitive availability for specific Linux kernel versions; and (iii) retrospective understanding of strategic primitive synthesis at different sophistication levels. This gap motivates three research questions: \textit{(1) How to classify and model the primitives in the general exploitation stage? (2) How to accurately and reliably extract exploit primitives for memory corruption vulnerabilities in the Linux kernel? (3) How to synthesize and chain primitives automatically?}

In this paper, we reformulate the end-to-end code generation problem into a primitive path optimization problem, iteratively discovering, validating, and synthesizing known and previously unknown primitives until finding solvable path and capable program states, merging the hard-coded patterns of known exploit primitives and the unshaped code paths into agentic primitives in concrete validation environments to reach reproducible and composable confirmation in the agentic loops.
To do so, we propose a tool-augmented multi-agent framework named \textsc{PrimSynth} to automatically discover, validate, synthesize, and chain exploit primitives. 
The framework couples five executable units: (i) vulnerability-directed execution based on directed fuzzing; 
(ii) LangChain-based agent orchestration for  \textit{plan $\rightarrow$ tool selection $\rightarrow$ action $\rightarrow$ critique $\rightarrow$ feedback} cycles; (iii) MCP-accessible tool adapters for observation, validation, and synthesis; (iv) a knowledge base that stores structural vulnerability facts for primitive analysis based on the knowledge graph; and (v) a rebootable validation environment built from QEMU and instrumented kernel images for specific target versions.

\textsc{PrimSynth} is extensively evaluated on 16 real-world Linux kernel CVE vulnerabilities, 
the results demonstrate a strategy synthesis rate (SSR) of 82.4\% with public PoC guidance and 61.3\% without, achieving 77.1\% overall with an average runtime of 72.4 seconds.
These results have three key implications for automated kernel exploitation research. First, the strong performance under incomplete information validates that \textsc{PrimSynth}'s iterative validation feedback mechanism effectively compensates for the absence of prior guidance, enabling practical zero-day analysis. Second, the interactive timescale (under 75 seconds) positions primitive synthesis as a usable tool for real-time security assessment workflows. Third, the successful synthesis of multi-step primitive chains across diverse vulnerability classes demonstrates that agentic approaches combining LLM-based reasoning with dynamic validation can achieve practical performance on vulnerability exploitability evaluation. 

The main contributions of this work are as follows:
\begin{itemize}
\item {We propose a systematic characterization (modeling, classification, and formalization) of exploit primitives, grounded by their identical capability or strategy pattern, which facilitates agent-based automatic primitive identification and synthesis.}

\item {We develop an automated method for identifying and validating primitives based on vulnerability-directed execution and a rebootable validation environment, which improves candidate objects grounding and accumulated primitive reproduction by path exploration under incomplete information. }

\item {We design and implement \textsc{PrimSynth}, a tool-augmented multi-agent framework with iterative feedback loops to discover, validate, and synthesize multi-primitive paths beyond single-class primitive settings. An anonymous artifact package is available at: \url{https://anonymous.4open.science/r/PrimSynth-8B7A}}.

\item {We evaluate \textsc{PrimSynth} on 16 real-world Linux kernel CVE vulnerabilities across 5 types. Results show that \textsc{PrimSynth} achieves reliable primitive extraction, maintaining 100\% primitive match
rate (PMR). For primitive synthesis, \textsc{PrimSynth} successfully synthesizes multi-primitive exploitation chains with 82.4\% SSR when the public PoC is available and a 61.3\% SSR without public PoC guidance. As for efficiency, \textsc{PrimSynth} completes end-to-end primitive synthesis in 72.4s on average, significantly outperforming semi-automated analysis or manual analysis that require hours to days. }
\end{itemize}


\section{Background and Motivation}

\textbf{Exploit Primitive Definition. }
In 2018, Branco \textit{et al} formalized the definition of \textit{\textbf{exploit primitive (EP)}} as an attack capability that an attacker can potentially achieve from a security vulnerability\cite{mathematicalmodeling}. They defined five major primitive properties in this model: \textit{arbitrary addresses}, \textit{arbitrary content}, \textit{arbitrary operation},\textit{ arbitrary number of times}, and \textit{at arbitrary time}. However, such definitions are too loose to implement in real-world applications. Google Project Zero \cite{ios-exp} later refined this definition in 2020, emphasizing the generality aspect of an exploit primitive: ``A capability granted during an exploit that is reasonably generic.'' Accordingly, the \textit{\textbf{exploit strategy}} is the low-level, vulnerability-specific method used to turn the vulnerability into a useful exploit primitive, and the corresponding \textit{\textbf{exploit technique}} is a reusable and reasonably generic strategy for turning one exploit primitive into another (usually more useful) exploit primitive.
Unlike userland exploitation that targets a single process, kernel exploitation breaks the Ring 3/Ring 0 isolation barrier to gain system-wide access. Characterizing primitive impacts remains challenging even with validation support.


\textbf{Prior Primitive Studies and Their Limitations. }
A rich body of work has studied exploitation primitives that form the building blocks of kernel exploits. On the target object identification front, ELOISE~\cite{chen2020systematic} systematically studied elastic objects as versatile exploitation targets, AlphaEXP~\cite{alphaexp} developed an expert system for identifying security-sensitive kernel objects, SCAVY~\cite{scavy} automated the discovery of memory corruption targets for privilege escalation, and Liu \textit{et al}.~\cite{thanos,puppet} studied freed object reuse and arbitrary write primitives via puppet objects. On the heap manipulation front, SLAKE~\cite{chen2019slake} provided systematic techniques for slab manipulation, K(H)eaps~\cite{kheaps} studied exploit reliability, DirtyCred~\cite{lin2022dirtycred} proposed credential swapping for privilege escalation, and RetSpill~\cite{zeng2023retspill} demonstrated igniting user-controlled data on the kernel stack. Maar\textit{ et al}.~\cite{def-amplified} demonstrated reliable exploits via defense-amplified TLB side-channel leaks.


\textbf{Motivation}. Although prior work has made substantial progress on individual exploitation techniques, these studies still leave a fundamental gap for exploitability analysis: exploit primitives are usually discovered, validated, and synthesized in isolation, and the resulting methods are tightly coupled to a specific vulnerability class, subsystem, or proof-of-concept. In practice, kernel exploitation is a chain of dependent capabilities. A tester must first localize a reachable vulnerability-triggering path, then identify a realizable primitive under the target environment, and finally compose multiple primitives into a valid exploitation sequence. However, existing approaches rarely provide a unified way to model primitive capability, environment prerequisites, temporal constraints, and validation evidence in one framework. As a result, they can often indicate whether a bug is exploitable in principle, but they cannot reliably answer which primitive is available, how it can be validated, or how multiple primitives can be composed across different kernel versions and execution contexts. This limitation is especially severe for memory-corruption vulnerabilities, where heap layout, object lifetime, race timing, and mitigation state all affect feasibility. Motivated by this gap, we develop \textsc{PrimSynth}, a framework that automatically discovers candidate primitives, validates them against concrete execution evidence, and synthesizes reusable primitive chains under both PoC-guided and PoC-free settings.


\section{Exploit Primitive Characterization}

\begin{table*}[htbp]
\centering
\scriptsize
\setlength{\tabcolsep}{0.3cm}
\caption{Formal Representation of Established Primitive Class}
\label{tab:primitive_identification}
\begin{tabular}{|m{2.4cm}|m{10.7cm}|m{3cm}|}
\hline
\textbf{Primitive Types} & \textbf{Formal Representation} & \textbf{Identical Object} \\
\hline
Out-of-Bounds &
$\exists \, \mathit{alloc}(s, \mathit{size}) \land \mathit{UserControllable}(\mathit{size}) \land \mathit{Reachable}(\mathit{alloc}, \mathit{syscall}) \land \mathit{StructType}(s, T)$ &
\parbox[t]{3cm}{Elastic Object \cite{chen2020systematic}} \\
\hline
Use-After-Free &
$\exists \, \mathit{alloc}(s) \land \exists \, \mathit{free}(p) \land \mathit{Alias}(p, s) \land \exists \, \mathit{use}(p') \land \mathit{Alias}(p', s) \land \mathit{TemporalOrder}(\mathit{free}, \mathit{use}) \land \mathit{Reachable}(\mathit{alloc}, \mathit{syscall}) \land \mathit{Reachable}(\mathit{use}, \mathit{syscall})$ &
\parbox[t]{3cm}{Freed Object \cite{thanos,puppet}} \\
\hline
Double-Free &
$\exists \, \mathit{alloc}(s) \land \exists \, \mathit{free}_1(p_1) \land \exists \, \mathit{free}_2(p_2) \land \mathit{Alias}(p_1, s) \land \mathit{Alias}(p_2, s) \land \mathit{free}_1(p_1) \neq \mathit{free}_2(p_2) \land \mathit{Reachable}(\mathit{free}_1(p_1), \mathit{syscall}) \land \mathit{Reachable}(\mathit{free}_2(p_2), \mathit{syscall})$ &
\parbox[t]{3cm}{Multiply-Freed Object \cite{thanos}} \\
\hline
Arbitrary-Write &
$\exists \, \mathit{alloc}(s) \land \exists \, \mathit{copy}(\mathit{to}, \mathit{from}, \mathit{len}) \land (\mathit{PointsTo}(\mathit{to}, s) \lor \mathit{FromField}(\mathit{to}, s, \mathit{offset})) \land (\mathit{FromField}(\mathit{len}, s, \mathit{offset}_l) \lor \mathit{FromField}(\mathit{to}, s, \mathit{offset}_t)) \land \mathit{Reachable}(\mathit{copy}, \mathit{syscall})$ &
\parbox[t]{3.6cm}{Bridge/Router Object  \cite{bridgerouter,puppet}} \\
\hline
Control-Flow-Hijacking &
$\exists \, \mathit{alloc}(s) \land \exists \, \mathit{call}(fptr) \land \mathit{FromField}(fptr, s, \mathit{offset}) \land \mathit{IsFunctionPtr}(fptr) \land \mathit{Reachable}(\mathit{alloc}, \mathit{syscall}) \land \mathit{Reachable}(\mathit{call}, \mathit{syscall}) \land \mathit{ControllableParams}(\mathit{call}, \geq 2)$ &
\parbox[t]{3cm}{Puppet Object \cite{puppet} }\\
\hline
Info-Leak &
$\exists \, \mathit{alloc}(s) \land \exists \, \mathit{leak}(\mathit{src}, \mathit{dst}) \land (\mathit{PointsTo}(\mathit{src}, s) \lor \mathit{FromField}(\mathit{src}, s, \mathit{offset})) \land \mathit{Reachable}(\mathit{alloc}, \mathit{syscall}) \land \mathit{Reachable}(\mathit{leak}, \mathit{syscall})$ &
\parbox[t]{3cm}{Disclosure Object  \cite{kleak} }\\
\hline
\end{tabular}
\end{table*}

\subsection{Formal Representation of Exploit Primitive}
We use exploit primitives as intermediate abstractions between bug triggers and end-to-end exploits.
To identify primitives through their \textit{identical objects} or \textit{gadget patterns} which constitute the general capabilities, we propose a systematic framework. Each primitive type exhibits distinct characteristics that can be statically identified through program analysis techniques provided in related works~\cite{kepler,thanos,bridgerouter}.

We use the following notation: the predicate $\mathit{alloc}(s, \mathit{size})$ denotes an allocation operation creating object $s$ with size $\mathit{size}$, while $\mathit{free}(p)$ denotes a de-allocation operation on pointer $p$. The relation $\mathit{Alias}(p_1, p_2)$ indicates that pointers $p_1$ and $p_2$ alias the same memory object, and $\mathit{PointsTo}(p, s)$ indicates that pointer $p$ points to object $s$. The predicate $\mathit{FromField}(v, s, \mathit{offset})$ indicates that value $v$ originates from the field at $\mathit{offset}$ in object $s$. The predicate $\mathit{Reachable}(op, \mathit{syscall})$ indicates that operation $op$ is reachable from a system call entry point, while $\mathit{TemporalOrder}(op_1, op_2)$ indicates that operation $op_1$ occurs before $op_2$ in execution order. Finally, $\mathit{UserControllable}(v)$ indicates that value $v$ is derived from user input.

Building on existing sanitization and object analysis techniques, we classify six primitive types prevalent in Linux kernel exploitation: Out-of-Bounds (OOB), Use-After-Free (UAF), Double-Free, Arbitrary-Write, Control-Flow-Hijacking (CFH), and Info-Leak. Table~\ref{tab:primitive_identification} presents their formal representations as logical predicates over memory operations, control flow, and data dependencies.

\subsection{Extended Exploit Strategy Representation}

Table~\ref{tab:primitive_identification} defines the six primitive types as logical predicates (the \emph{capability view}).
For primitive reproduction, we further define a strategy contract (the \emph{execution view}) that binds those predicates to step-wise constraints, validation signals, and code-generation actions.
A strategy is represented as:
\[
\mathcal{S}=\langle \mathcal{P},\mathcal{B},\mathcal{K},\mathcal{V},\mathcal{G},\mathcal{A}\rangle .
\]
Where, $\mathcal{P}=[p_1,\ldots,p_n]$ is an ordered primitive chain; $\mathcal{B}=[\Psi_1,\ldots,\Psi_{n-1}]$ is a set of \textit{\textbf{primitive path code synthesis rules}}, where each $\Psi_i$ defines how to generate executable path code snippets between consecutive primitives grounded in candidate objects from $\mathcal{K}$; $\mathcal{K}$ is the discovery-side knowledge tuple (entry syscalls, reachable paths, object priors, and environment gating); $\mathcal{V}$ is the validation oracle tuple (human-friendly observable outcomes and machine-readable validation signals); $\mathcal{G}$ is the synthesis tuple (step-level facts and code templates/strategy constraints); and $\mathcal{A}$ is the final exploit artifact tuple (generated code plus evidence log).

For each primitive transition $p_i \rightarrow p_{i+1}$, a primitive path code synthesis rule is defined as :
\[
\Psi_i=\Psi_i^{obj}\land\Psi_i^{time}\land\Psi_i^{env}\land\Psi_i^{sig},
\]

where the \textit{object constraint} $\Psi_i^{obj}$ specifies which candidate objects (e.g., \texttt{msg\_msg}, \texttt{pipe\_buffer}) and their field offsets are required in the synthesized code;
the \textit{temporal constraint} $\Psi_i^{time}$ defines the allocation/free/reuse sequence timing required for heap manipulation primitives;
the \textit{environment constraint} $\Psi_i^{env}$ captures kernel version, mitigation status, and runtime prerequisites for code execution;
the \textit{validation constraint} $\Psi_i^{sig}$ enforces minimum signals that confirm the synthesized code achieves the expected primitive capability.
Code generation proceeds only when $\Psi_i$ is satisfied, ensuring each synthesized step is grounded in validated candidate objects rather than abstract transitions. 


Each extracted primitive is validated through machine-observable \textit{\textbf{validation signals}} (will discuss in detail in Section \ref{sec:validation}) generated by primitive identification, debugging, or compilation tools. 
Among them, we use struct layouts and field offsets provided by the static tool PODE \cite{puppet} to select objects; Then, we confirm actual memory corruption by semantics and offsets provided by memory error reports such as KASAN;
For timing control, we track the complete object lifecycle through debugger watchpoints to verify heap manipulation sequences; For path reachability, we capture and analyze the execution trace from syscall entry to vulnerability location, and check the security boundary conditions;
Finally, to confirm success, we detect intermediate achievements and verify final exploitation outcomes. 
If validation fails, the system adapts: mismatched object offsets trigger re-hypothesization, failed timing sequences lead to parameter adjustment, unreachable paths require entry point modification, and missing achievement signals necessitate chain redesign. 


\subsection{Representation Mappings for Agent Pipeline}
To connect strategy representation with the agent pipeline, we define three mappings that ground code path synthesis in concrete validation and generation:

\textbf{Discovery mapping} $\mathcal{M}_D: q \mapsto \mathcal{K}$, populates the knowledge tuple $\mathcal{K}$ with candidate objects (e.g., \texttt{msg\_msg}, \texttt{pipe\_buffer}), step hypotheses with reference primitives, and validation signals for primitive path code synthesis rules;

\textbf{Validation mapping} $\mathcal{M}_V: (\mathcal{P},\mathcal{B},\mathcal{K}) \mapsto \mathcal{V}$, verifies that primitive path code synthesis rules ($\Psi_i \in \mathcal{B}$) are satisfied by concrete candidate objects through validation signals (S1--S9), producing the validation oracle tuple $\mathcal{V}$;

\textbf{Synthesis mapping} $\mathcal{M}_S: (\mathcal{P},\mathcal{B},\mathcal{K},\mathcal{V}) \mapsto (\mathcal{G},\mathcal{A})$, instantiates validated primitive path code synthesis rules into executable code patterns stored in $\mathcal{G}$, generating the artifact tuple $\mathcal{A}$ containing synthesized exploit code grounded in validated candidate objects.

We separate reusable strategy skeletons from environment-specific bindings.
The reusable part is captured in $(\mathcal{P},\mathcal{B},\mathcal{G})$, while kernel-version and subsystem-dependent factors are isolated in environmental primitive path code synthesis rules (i.e., $\Psi_i^{env}$) and re-checked by $\mathcal{M}_V$ for each case.
Hence, 
the same strategy can be adapted across Linux versions/subsystems with re-validation of environment-specific constraints

\section{Design of \textsc{PrimSynth} Framework}

\begin{figure*}
\centering
  \includegraphics[width=\textwidth]{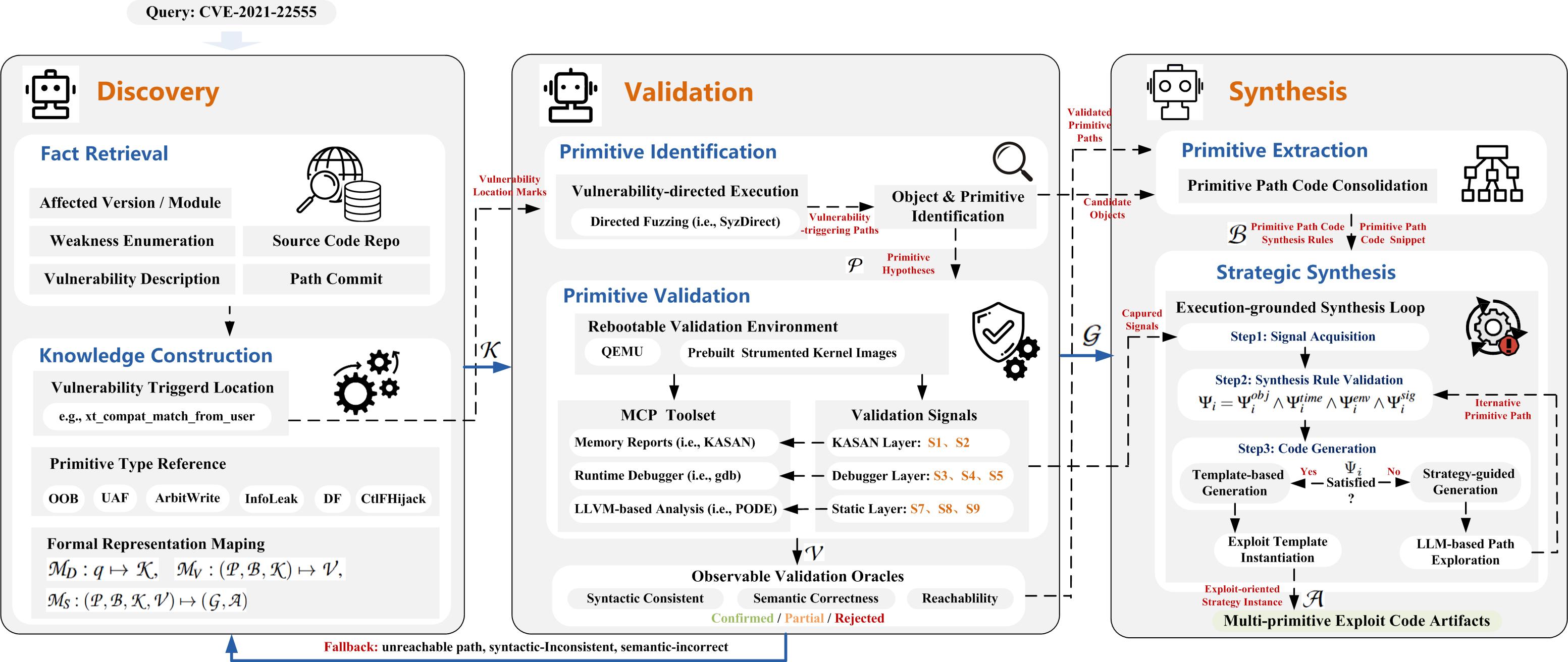}
  \caption{Overview workflow of \textsc{PrimSynth} framework}
  \label{fig:overview}
\end{figure*}

\subsection{Overview}

\textsc{PrimSynth} is a unified, agent-driven framework for reproducible primitive analysis and synthesis. As Figure~\ref{fig:overview} shows, 
the framework executes a three-stage agent pipeline---Discovery, Validation, and Synthesis---and closes the loop through structured validation feedback.

The \textbf{Discovery} agent collects facts and contexts. It performs \textit{fact retrieval} and \textit{knowledge construction} to collect affected software/version, weakness type, patch reference, vulnerability locations, primitive-type priors, and reproducibility context. The reproducibility context usually contains claimed reproducible PoC code, vulnerability-triggering input, or a demonstration image with a vulnerable version of the Linux kernel.

The \textbf{Validation} agent confirms feasibility and gathers evidence. It first performs \textit{primitive identification} using vulnerability-directed execution to derive the \textit{vulnerability-triggering path} and identify candidate objects with primitive hypotheses along the vulnerability-triggering path. Then, it performs \textit{primitive validation} to verify semantic correctness and reproducibility of the primitive hypotheses with tool-assisted execution. 

The \textbf{Synthesis} agent composes validated primitives into exploit code. It first performs \textit{primitive extraction} to extract primitive path code snippets from the candidate objects. Then it leverages \textit{strategic synthesis} to compose path code snippets into exploit code artifacts based on the primitive path code synthesis rules.


The agent collaboration loop is organized as \emph{plan $\rightarrow$ tool selection $\rightarrow$ action $\rightarrow$ critique $\rightarrow$ fallback}.
Given a query (typically a CVE ID), the Discovery Agent first materializes \textit{Vulnerability Location Marks} from the procedure of fact retrieval and knowledge construction. Then the Validation Agent schedules identification and validation actions to derive candidate objects and primitive paths. After that, the Synthesis Agent uses the procedure of primitive extraction to consolidate the validated primitive paths into code snippets, and strategic synthesis composes the primitive path code snippets to exploit code artifacts.
The core procedure is abstracted as \textbf{\textit{vulnerability location mark}s $\rightarrow$ \textit{vulnerability-triggering path} $\rightarrow$ \textit{candidate objects} $\rightarrow$ \textit{primitive hypotheses} $\rightarrow$ \textit{validated primitive paths} $\rightarrow$ \textit{primitive path code snippet} $\rightarrow$ \textit{multi-primitive exploit code}}.
The loop maintains iteration-level state observables and failure-inducing feedback (e.g., unreachable path, invalid strategy, or weak observable signal), which guides targeted recovery instead of restarting from scratch. This closed-loop design ensures that primitive hypotheses are accepted only if they satisfy both structural plausibility and execution-level evidence. 

\subsection{Factual Knowledge Discovery}

The Discovery Agent operates through a structured two-phase procedure to transform raw vulnerability reports into actionable knowledge structures.

\textit{1) Fact Retrieval}. The Discovery Agent employs hybrid information extraction techniques combining structured parsing with semantic web search. It systematically queries security databases (CVE, NVD, security advisories), version control repositories, and patch archives to extract affected software identities, version ranges, CWE classifications, and reference materials including official patches, commit diffs, and vendor security bulletins. A keyword-based web retrieval in this phase ensures comprehensive coverage of vulnerability artifacts.

\textit{2) Knowledge Construction}. The Discovery Agent synthesizes retrieved facts into a unified knowledge graph representation tailored for downstream primitive analysis. This process involves entity resolution to normalize software names and version notations, relation extraction to establish dependencies between vulnerability locations and affected components, and schema alignment to map unstructured facts (PoC snippets, crash reports, procedural illustration) into structured reproducibility context. The constructed knowledge base encodes vulnerability-triggering conditions, entry point locations in source code, and primitive-type priors derived from historical patterns, forming a queryable foundation that enables the Validation and Synthesis agents to perform targeted primitive identification and strategic composition.

Concretely, the Discovery Agent outputs a normalized case profile including: (i) vulnerability facts (affected range, weakness class, patch references), (ii) candidate entry syscalls and vulnerable locations, (iii) identified step-wise primitives, and (iv) environment-gating requirements.
These outputs are scored with confidence tags, so downstream components can distinguish high-certainty anchors from speculative hypotheses.
When the Validation agent reports a contradiction (e.g., path mismatch or missing validation signal), the Discovery agent re-ranks candidates and updates the profile without discarding confirmed facts.

\subsection{Primitive Identification and Validation}
\label{sec:validation}
The Validation Agent first identifies the candidate objects along the vulnerability-triggering path, then it collects the corresponding primitive hypotheses and verifies their semantic/syhtactic correctness and reproducibility with tool-assisted execution to derive validated primitive paths.


\textit{1) Primitive Identification:} the Validation Agent consumes vulnerability location marks and 
generates directed fuzzing campaigns to correctly activate the expected vulnerability-triggering path.
It first instruments the vulnerability location marks as the objective of directed fuzzing campaigns, then enumerates primitive hypotheses by identifying and grouping the prerequisite operations on reached objects. These operations include allocation/free/use relations and function-pointer invocation.
After that, the agent identifies the candidate objects along the vulnerability-triggering path and collects the primitive hypotheses for these objects.
The output of this procedure is a ranked set of primitive hypotheses, candidate objects, and expected side effects.

\textit{2) Primitive Validation:} the agent evaluates each primitive hypothesis for semantic/syhtactic correctness and path reachability through machine-observable validation signals generated by primitive identification, debugging, or compilation tools. Specifically, including the following 9 validation signals across three layers:
\begin{itemize}
  \item \textit{KASAN Layer}: Memory error reports providing corruption semantic and offsets (\textbf{S1}) and execution trace from syscall entry to vulnerability location(\textbf{S2});
  \item \textit{Debugger Layer}: Runtime state observation via dbg-tool capturing complete object lifecycle (\textbf{S3}), security boundary conditions(\textbf{S4}), intermediate achievement (\textbf{S5}), and final exploitation outcome (\textbf{S6});
  \item \textit{Static Layer}: LLVM-based analysis via PODE \cite{puppet} providing object layout and field offsets (\textbf{S7}), alias relation (\textbf{S8}), and pointer instruction recognition (\textbf{S9}).
\end{itemize}

These validation signals are aggregated into observable outcomes for the user interface, including:
\begin{itemize}
\item \textit{confirmed} (all primitive path code synthesis rules $\Psi_i$ satisfied by candidate objects);
\item \textit{partial} (some $\Psi_i^{sig}$ met but $\Psi_i^{obj}$ or $\Psi_i^{time}$ pending);
\item\textit{rejected} (signal mismatches candidate objects detected).
\end{itemize}

It is noteworthy that execution trace (S2) and static analysis results (S7--S9) primarily serve as synthesis inputs that feed into the Synthesis Agent's primitive extraction and upgrading strategy selection.
Machine-readable failure reasons directly drive fallback policies: unreachable paths trigger Discovery-side path revision, while signal mismatches trigger Synthesis-side strategy rewrites.
If current primitive hypotheses are unreachable, syntactically inconsistent, or semantically incorrect in the validation environment, the loop falls back to knowledge discovery and revises the reproducibility context while preserving already validated evidence. The output of this procedure is a set of validated primitive paths.


\subsection{Primitive Extraction and Synthesis}
\label{sec:sythesis}
After validation, the Synthesis Agent extracts primitive path code snippets from the candidate objects with the  validated primitive paths. Then it leverages \textit{strategic synthesis} to compose path code snippets into multi-primitive exploit code artifacts based on the primitive path code synthesis rules.


\subsubsection{Primitive Extraction} consolidates confirmed operation slices into reusable primitive path code snippets with primitive path code synthesis rules ($\Psi_i$) governing how to compose executable code between consecutive steps.
This step transforms noisy traces into a compact strategy-ready representation and removes transitions unsupported by validation signals.

\subsubsection{Strategic Synthesis} composes primitive path code snippets through an \textit{\textbf{execution-grounded synthesis loop}}.

\textbf{Step 1: Signal Acquisition}: KASAN reports capture S1 (memory corruption), execution traces capture S2 (path hits i.e., control flow reachability as the syscall trace leading to the crash), and dbg-tool watchpoints capture S3 (object lifecycle).
The dbg-tool framework enables precise observation through memory watchpoints, object lifecycle tracking, and crash evidence capture, providing the validation signals required for confirming primitive path code synthesis rules.

\textbf{Step 2: Synthesis Rule Validation}: each primitive path must satisfy primitive path code synthesis rules $\Psi_i = \Psi_i^{obj} \land \Psi_i^{time} \land \Psi_i^{env} \land \Psi_i^{sig}$ grounded in candidate objects. For example, a Step 1 OOB rule $\Psi_1$ requires S2 signals confirming corruption of candidate object \texttt{msg\_msg} headers ($\Psi_1^{obj}$ satisfied), while a Step 2 Info-Leak rule $\Psi_2$ requires S4 signals confirming disclosure through \texttt{msgrcv()} paths ($\Psi_2^{sig}$ satisfied).

\textbf{Step 3: Code Generation}: upon $\Psi_i$ satisfaction (all object, temporal, environment, and validation constraints met), template-based instantiation reuses code patterns from $\mathcal{G}$ bound to validated primitive path and candidate objects; if signals indicate under-specified $\Psi_i^{obj}$ (unknown candidate object fields), strategy-guided path exploration via LLM generates alternative primitive paths targeting different candidate objects (e.g., \texttt{sk\_buff} replacing \texttt{msg\_msg}) for iterative validation.

Generated code artifacts are re-injected into the Validation Agent, forming an execution-grounded synthesis loop.


\subsubsection{Signal-driven Strategy Selection}
The Synthesis Agent implements a dynamic policy that selects between template-based and strategy-guided generation based on validation signals captured by tools, with S7--S9 serving as object-grounded constraints that determine primitive upgrading feasibility.
The upgrading strategies (e.g., OOB~$\rightarrow$~InfoLeak~$\rightarrow$~ArbitraryWrite) are selected and instantiated based on S7--S9 signals that identify which object transitions are structurally feasible.

\textbf{Template-based generation}: When KASAN reports confirm both S1 signals (memory corruption at expected candidate object offsets) and S2 signals (syscall trace hitting vulnerability locations), and dbg-tool watchpoints capture S3 signals (allocation-free-reuse matching candidate object assumptions), the system performs \textbf{template instantiation} from $\mathcal{G}$. This path requires all primitive path code synthesis rules $\Psi_i = \Psi_i^{obj} \land \Psi_i^{time} \land \Psi_i^{env} \land \Psi_i^{sig}$ to be satisfied by validation signals on known candidate objects.

\textbf{Strategy-guided Generation}: When GDB observations reveal incomplete S3 signals (missing candidate object lifecycle events), or KASAN reports show inconsistent S1 patterns (corruption at unexpected candidate object offsets) or incomplete S2 traces (unreachable paths for candidate objects), the Synthesis Agent falls back to LLM-based path exploration to generate iterative primitive paths to different candidate objects. This procedure iteratively discovers new candidate objects and accumulates signals---from initial S2 detection on experimental objects through S3 state establishment to S5 completion---until sufficient $\Psi_i^{sig}$ thresholds are met and new primitive path code synthesis rules can be promoted into $\mathcal{G}$.

The Synthesis Agent monitors $\Psi_i^{sig}$ at each step transition, automatically decides a strategy when signal mismatches on candidate objects occur (e.g., KASAN offset deviations on \texttt{msg\_msg}, dbg-tool allocation misses for \texttt{pipe\_buffer}, trace deviations), maintaining synthesis robustness across diverse vulnerability contexts.

\section{Implementation}

We implement \textsc{PrimSynth} as a tool-augmented multi-agent system with unified callable interfaces and persistent case-level states.
The implementation couples five executable layers:
\textbf{(i)} vulnerability-directed execution based on directed fuzzing;
\textbf{(ii)} LangChain-based agent orchestration; 
\textbf{(iii)} MCP-accessible tool adapters for observation, validation, and synthesis; 
\textbf{(iv)} a knowledge base that stores structural vulnerability facts based on the knowledge graph. 
and \textbf{(v)} a rebootable validation environment built from QEMU and instrumented kernel images. 

\textbf{Vulnerability-directed Execution}
Runtime execution is implemented based on directed fuzzing (i.e.,
SyzDirect \cite{tan2023syzdirect}). It takes vulnerability location marks into instrumentation points and mutation objectives, then schedules directed campaigns to activate the vulnerability-triggering path validation.
The system prioritizes location-directed fuzzing campaigns to attain accurate path triggering.

\textbf{MCP-based Toolset.} All analyzers and validators are exposed as MCP-style tools, including knowledge access, data extraction, validation, and synthesis interfaces, and can be attached from either in-environment or external runtimes. The current tools capture validation signals while keeping the MCP interface open for additional observers and validators.
In our implementation, orchestration supports timeout/retry policy, confidence-aware tool selection, and signal-based aggregation for final step decisions.

\textbf{Knowledge Base.} The knowledge base stores both static and dynamic evidence based on Neo4j.
Static entries include vulnerability facts, patch diffs, coarse-grained vulnerability locations, reusable strategy representations, and primitive path code synthesis rules; dynamic entries include runtime observables, validation outcomes, and bridging strategies extracted from prior iterations.
To support cross-case transfer, strategy tuples are indexed by primitive-centric step pattern and object/temporal/environment signatures.
During synthesis, template lookup from $\mathcal{G}$ is jointly conditioned on primitive-centric step, object constraints, and signal-based evidence.
Selected code templates or strategy prompts are materialized as versioned generation context so intermediate synthesis outputs can be reproduced, compared, and rolled back iteratively.

\textbf{Rebootable Validation Environment.} The rebootable environment is built with QEMU and prebuilt instrumented kernel images for specific target versions, using deterministic boot parameters and controllable mitigation settings.
Each iteration begins from a clean snapshot and records execution artifacts (kernel logs, KASAN reports, trace events, and debugger checkpoints).
On failure, the environment can be reset and replayed with adjusted strategy parameters while preserving iteration history and evidence memory.

This rebootable design is also used during synthesis for single- and multiple-primitive validation: intermediate synthesized PoC code are repeatedly replayed in the same kernel image to verify that validation signals confirm primitive path code synthesis rules ($\Psi_i$) are satisfied by target candidate objects before transfer into later steps.


\section{Evaluation}

To evaluate \textsc{PrimSynth}, we conduct experiments on the automated discovery, validation, and synthesis of exploit primitives, aiming to address four research questions:

 \textbf{RQ1 (Primitive Extraction Accuracy)}: How effective is the primitive identification and extraction method?

 \textbf{RQ2 (Synthesis Utility)}: Can \textsc{PrimSynth} synthesize the initial primitive and upgraded primitives into executable and semantically-correct code?

 \textbf{RQ3 (Overall Effectiveness)}: How does \textsc{PrimSynth} compare with baseline tools in end-to-end multi-primitive reproduction?

 \textbf{RQ4 (Efficiency and LLM Performance)}: What is the runtime efficiency of the multi-agent framework based on different tools and LLMs?

\begin{table*}[htbp]
\centering
\scriptsize
\caption{CVE Benchmark used in the evaluation\protect\footnotemark}
\label{tab:cve-overview}

\begin{tabular}{|p{1.8cm}|p{1.2cm}|p{3.5cm}|p{0.7cm}|p{3cm}|p{0.7cm}|p{4cm}|}
\hline
\textcolor{black}{\textbf{CVE ID}} & \textcolor{black}{\textbf{Vuln. Type}} & \textcolor{black}{\textbf{Vuln. Location}} & \textcolor{black}{\textbf{EPSS}} & \textcolor{black}{\textbf{Primitive Ref.}} & \textcolor{black}{\textbf{PoCs}} & \textcolor{black}{\textbf{Env. Conditions}} \\
\hline
CVE-2022-2588 & UAF & \texttt{struct cls\_route\_filter} & 0.54 & UAF$\rightarrow$DF$\rightarrow$ArbWrite & \cmark & \texttt{CONFIG\_NET\_SCHED}, 2+ CPUs \\
CVE-2022-0995 & OOB & \texttt{struct watch\_queue} & 0.28 & OOB$\rightarrow$InfoLeak$\rightarrow$ArbWrite$\newline \rightarrow$CFH & \cmark & \texttt{CONFIG\_WATCH\_QUEUE}, KASLR on \\
CVE-2022-1043 & UAF & \texttt{struct io\_uring} & 0.19 & UAF$\rightarrow$InfoLeak$\rightarrow$ArbWrite & \cmark & \texttt{CONFIG\_IO\_URING}, KASLR on \\
CVE-2019-15666 & OOB & \texttt{struct xfrm\_policy} & 0.05 & OOB$\rightarrow$UAF$\rightarrow$ArbWrite & \cmark & \texttt{CONFIG\_XFRM\_MIGRATE} \\
CVE-2022-32250 & UAF &  \texttt{struct nft\_set\_elem} & 0.02 & UAF$\rightarrow$InfoLeak$\rightarrow$ArbWrite  & \cmark & \texttt{CONFIG\_NF\_TABLES} \\
CVE-2024-53197 & OOB & \texttt{struct usb\_host\_config} & 0.02 & OOB & \xmark & \texttt{CONFIG\_USB}, SMEP/SMAP off \\
CVE-2022-0185 & IntOF$\rightarrow$OOB & \texttt{struct fs\_parameter} & 0.02 & IntOF$\rightarrow$OOB$\rightarrow$InfoLeak$\newline \rightarrow$ArbWrite & \cmark & \texttt{CONFIG\_FS\_CONTEXT} \\
CVE-2022-1015 & IntOF & \texttt{struct nf\_tables} & 0.01 & \makecell[l]{IntOF$\rightarrow$OOB$\rightarrow$InfoLeak \\ $\rightarrow$CFH} & \cmark & \texttt{CONFIG\_NF\_TABLES}, libmnl \\
CVE-2022-2639 & IntOF &\texttt{struct sw\_flow\_actions} & 0.01 & IntOF & \xmark & \texttt{CONFIG\_OPENVSWITCH} \\
CVE-2021-42008 & OOB & \texttt{struct sixpack} & 0.01 & OOB$\rightarrow$InfoLeak$\rightarrow$CFH & \cmark & \texttt{CONFIG\_SIXPACK}, SMEP/SMAP off \\
CVE-2022-27666 & OOB & \texttt{struct skcipher\_walk} & 0.01 & OOB$\rightarrow$InfoLeak$\rightarrow$ArbWrite  & \cmark & \texttt{CONFIG\_XFRM\_ESP},FUSE for PF \\
CVE-2022-25636 & OOB & \texttt{struct flow\_rule} & 0.005 & OOB$\rightarrow$UAF$\rightarrow$CFH & \cmark & \texttt{CONFIG\_NET\_SCHED} \\
CVE-2021-22600 & DF & \texttt{struct packet\_ring\_buffer} & 0.002 & UAF$\rightarrow$DF$\rightarrow$InfoLeak$\rightarrow$CFH & \cmark & \texttt{CONFIG\_NF\_TABLES}\\
CVE-2025-38477 & Race$\rightarrow$UAF & \texttt{struct qfq\_aggregate} & 0.00049 & Race$\rightarrow$UAF & \xmark & \texttt{CAP\_NET\_ADMIN} \\
CVE-2026-23427 & UAF & \texttt{struct ksmbd\_tree\_conn} & 0.00045 & UAF & \xmark & \texttt{CONFIG\_KSMBD\_SERVER} \\
CVE-2024-39486 & Race$\rightarrow$UAF & \texttt{struct inet6\_dev} & 0.00029 & Race$\rightarrow$UAF$\rightarrow$ArbWrite   & \cmark & \texttt{CONFIG\_IPV6\_MROUTE}, 2+ CPUs \\
\hline
\end{tabular}
\end{table*}
\footnotetext{EPSS scores are obtained from the FIRST.org Exploit Prediction Scoring System (EPSS) API \url{https://api.first.org/data/v1/epss}) on 2026-06-10.}

\textbf{Experimental Setup.} As Table \ref{tab:cve-overview} lists, to evaluate the performance of \textsc{PrimSynth}, we build a benchmark of 16 Linux kernel vulnerabilities with assigned CVE IDs across 5 vulnerability types: Out-of-Bounds (OOB), Use-After-Free (UAF), Double-free (DF), Race Condition (Race), and Integer Overflow (IntOverflow). 
To evaluate \textsc{PrimSynth} with a constrained information condition, 25\% of the vulnerabilities have no public exploitation description or executable PoC (marked as $\times$ in Table~\ref{tab:cve-overview}). 
Since no existing baseline covers the full PrimSynth workflow, we compare against prior work at the closest comparable subtask granularity.

\subsection{RQ1: Primitive Extraction Accuracy}

\subsubsection{Metrics and Setup}
We evaluate primitive extraction with two metrics:

\textbf{PMR (Primitive Match Rate)}: the percentage of cases whose extracted primitive with candidate object type matches ground truth \emph{and} whose grounded capabilities are validated by signals (e.g., UAF/OOB cases require corruption semantics and reachable paths via S1/S2).

\[PMR=\frac{|\{cases:type\_match\land capability\_evidence\}|}{|total\_cases|}\]

\textbf{OFLA (Object-Field Localization Accuracy)}: the precision of object-field localization under \texttt{FromField}/\texttt{StructType} constraints in Section 3, validated by static-dynamic alignment (S7 vs S1).

\[OFLA = 1 - \frac{|S7.offset - S1.offset|}{sizeof(struct)}\]

The \textbf{PMR} and \textbf{OFLA} pair together measure whether extracted primitives are both semantically correct and operationally grounded: PMR addresses class-level correctness, while OFLA validates object-field grounding required by identical-object-centric reasoning (OFLA $> 0.8$ indicates accurate localization).

\subsubsection{Results}

\begin{table*}[htbp]
\centering
\scriptsize
\caption{Primitive Extraction Accuracy: Matched Primitives and Object-Field Localization (OFLA)}
\label{tab:rq1-extraction-results}
\begin{tabular}{|l|c|c|p{6.5cm}|c|}
\hline
\textbf{CVE ID} & \textbf{Steps} & \textbf{PMR} & \textbf{Identical Objects in Matched Primitives} & \textbf{Avg OFLA} \\
\hline
CVE-2019-15666 (OOB$\rightarrow$UAF$\rightarrow$ArbWrite) & 3 & \cmark & xfrm\_policy, xfrm\_state & 0.83 \\
CVE-2021-22600 (DF$\rightarrow$InfoLeak$\rightarrow$CFH) & 4 & \cmark & packet\_ring\_buffer$\rightarrow$signalfd\_ctx$\rightarrow$passwd & 0.86 \\
CVE-2021-42008 (OOB$\rightarrow$InfoLeak$\rightarrow$CFH) & 3 & \cmark & sixpack$\rightarrow$userfaultfd$\rightarrow$modprobe\_path & 0.88 \\
CVE-2022-0185 (IntOF$\rightarrow$OOB$\rightarrow$InfoLeak$\rightarrow$ArbWrite) & 4 & \cmark & legacy\_fs\_ctx$\rightarrow$msg\_msg$\rightarrow$pipe\_buffer & 0.82 \\
CVE-2022-0995 (OOB$\rightarrow$InfoLeak$\rightarrow$ArbWrite$\rightarrow$CFH) & 4 & \cmark & watch\_queue$\rightarrow$msg\_msg$\rightarrow$pipe\_buffer $\rightarrow$modprobe\_path & 0.92 \\
CVE-2022-1015 (IntOF$\rightarrow$OOB$\rightarrow$InfoLeak$\rightarrow$CFH) & 4 & \cmark & nft\_expr$\rightarrow$msg\_msg$\rightarrow$modprobe\_path & 0.86 \\
CVE-2022-1043 (UAF$\rightarrow$InfoLeak$\rightarrow$ArbWrite) & 3 & \cmark & io\_uring$\rightarrow$cred$\rightarrow$suid & 0.84 \\
CVE-2022-2588 (UAF$\rightarrow$DF$\rightarrow$ArbWrite) & 3 & \cmark & route4\_filter$\rightarrow$file$\rightarrow$cred & 0.86 \\
CVE-2022-2639 (IntOF$\rightarrow$OOB$\rightarrow$UAF$\rightarrow$ArbWrite) & 5 & \cmark & sw\_flow\_actions$\rightarrow$msg\_msg$\rightarrow$pipe\_buffer & 0.81 \\
CVE-2022-25636 (OOB$\rightarrow$UAF$\rightarrow$CFH) & 3 & \cmark & flow\_rule$\rightarrow$msg\_msg$\rightarrow$ethtool\_ops/pipe\_buffer & 0.83 \\
CVE-2022-27666 (OOB$\rightarrow$InfoLeak$\rightarrow$ArbWrite) & 3 & \cmark & skcipher\_walk$\rightarrow$user\_key\_payload$\rightarrow$msg\_msg & 0.80 \\
CVE-2022-32250 (UAF$\rightarrow$InfoLeak$\rightarrow$ArbWrite) & 3 & \cmark & nft\_set\_elem$\rightarrow$user\_key\_payload$\rightarrow$io\_uring $\rightarrow$modprobe\_path & 0.78 \\
CVE-2024-39486 (Race$\rightarrow$UAF$\rightarrow$ArbWrite) & 3 & \cmark & inet6\_dev$\rightarrow$ifmcaddr6$\rightarrow$modprobe\_path & 0.68 \\
CVE-2024-53197 (OOB) & 1 & \cmark & usb\_host\_config & 0.75 \\
CVE-2025-38477 (Race$\rightarrow$UAF) & 2 & \cmark & qfq\_aggregate & 0.65 \\
CVE-2026-23427 (UAF$\rightarrow$ArbWrite$\rightarrow$CFH) & 3 & \cmark & ksmbd\_tree\_conn$\rightarrow$ksmbd\_file$\rightarrow$modprobe\_path & 0.79 \\
\hline
\multicolumn{2}{|l|}{\textbf{Total}} & \textbf{100\%} & \textbf{Avg OFLA (w/ offset data): 0.81} & \\
\hline
\end{tabular}
\end{table*}

Table~\ref{tab:rq1-extraction-results} presents the primitive extraction results for all 16 CVE cases, including the complete primitive chain with stage count, PMR, identical objects identified, and average OFLA. All 16 cases achieve 100\% PMR  with OFLA ranging from 65\% to 92\%.
The OFLA distribution reflects varying localization complexity: OOB vulnerabilities achieve 75--92\% through explicit structure field relationships; UAF vulnerabilities achieve 78--86\% through lifecycle signal correlation; Race-conditioned vulnerabilities exhibit 65--68\% due to non-deterministic execution.
The validation signal framework grounds primitive extraction in concrete execution evidence through S1--S9 signal chains.

Notably, the framework identifies valid primitive paths for recently disclosed vulnerabilities lacking public PoC documentation. For CVE-2024-53197 (USB Audio OOB, EPSS 0.02, no public exploit), the tool combination discovers the \texttt{usb\_host\_config} to \texttt{pipe\_buffer} exploitation chain through S7 object layout analysis, S2 trace validation, and S3 lifecycle tracking. For CVE-2024-39486 (Race-to-UAF in IPv6 MLD), SyzDirect's directed mutation generates candidate interleavings while PODE identifies \texttt{inet6\_dev} and \texttt{ifmcaddr6} as identical objects through RCU pointer dependency analysis.

\begin{table*}[htbp]
\centering
\scriptsize
\caption{Tool Combination Analysis: Candidate Object Discovery and Primitive Path Exploration}
\label{tab:rq1-tool-combination}
\begin{tabular}{|l|c|c|c|c|c|}
\hline
\textbf{CVE ID} & \textbf{Discovery} & \textbf{+SyzDirect} & \textbf{+PODE} & \textbf{Full Combo} & \textbf{Ground Truth} \\
\hline
CVE-2022-0995 (OOB) & 12/28 (43\%) & 22/28 (79\%) & 18/28 (64\%) & 27/28 (96\%) & 28 objects \\
CVE-2022-2588 (UAF) & 8/15 (53\%) & 11/15 (73\%) & 14/15 (93\%) & 15/15 (100\%) & 15 objects \\
CVE-2022-1043 (UAF) & 6/12 (50\%) & 9/12 (75\%) & 10/12 (83\%) & 12/12 (100\%) & 12 objects \\
CVE-2024-53197 (OOB) & 4/9 (44\%) & 6/9 (67\%) & 8/9 (89\%) & 9/9 (100\%) & 9 objects \\
CVE-2024-39486 (Race) & 3/8 (38\%) & 5/8 (63\%) & 6/8 (75\%) & 7/8 (88\%) & 8 objects \\
CVE-2026-23427 (UAF) & 5/11 (45\%) & 7/11 (64\%) & 9/11 (82\%) & 10/11 (91\%) & 11 objects \\
\hline
\textbf{Avg Recall} & \textbf{45\%} & \textbf{70\%} & \textbf{81\%} & \textbf{96\%} & -- \\
\textbf{Avg Iterations} & 18.5 & 12.3 & 8.7 & 4.2 & -- \\
\hline
\end{tabular}
\end{table*}

\textbf{Ablation study}. To further validate the systematic characterization’s effectiveness in discovering candidate objects and primitive paths,
we conduct an ablation study 
examining the contribution of SyzDirect fuzzing and PODE static analysis when combined with the Discovery agent knowledge construction. Table~\ref{tab:rq1-tool-combination}
presents the candidate object recall and primitive path coverage under four tool configurations across representative CVEs spanning OOB, UAF, Double-Free, and Race categories.
The experimental results reveal complementary strengths across tool combinations. Discovery Agent alone, relying on knowledge base retrieval and patch analysis, achieves only 45\% candidate recall with high iteration counts (18.5 average) due to candidate space explosion without runtime validation guidance. Adding SyzDirect directed fuzzing improves recall to 70\% by generating concrete execution traces that validate reachable syscall sequences. The full combination achieves 96\% average recall with only 4.2 iterations, demonstrating synergistic effects where Discovery provides initial hypotheses, SyzDirect validates execution feasibility, and PODE refines object field layouts.

\textbf{Answer to RQ1:} Systematic characterization achieves reliable primitive extraction by integrating knowledge construction, directed execution, and static analysis. This combination reduces candidate object space by 98.5\% (from 340 to 5.2 average candidates) while maintaining 100\% primitive match rate, enabling discovery of alternative exploitation paths beyond documented PoC techniques with OFLA ranging 0.65--0.92 across vulnerability classes.

\subsection{RQ2: Synthesis Utility}

\subsubsection{Metrics and Setup}
We design three metrics for primitive synthesis evaluation:

\textbf{SSR (Strategy Synthesis Rate)}: step-wise synthesis rate measuring code generation fidelity, quantifying how completely the synthesized code implements the extracted primitive path's step transitions according to strategy representation.

\[
\text{SSR} = \frac{1}{|\mathcal{B}|} \sum_{\Psi_i \in \mathcal{B}} \frac{|\{\psi \in \Psi_i \mid \text{code\_gen}(\psi) = \text{SAT}\}|}{|\Psi_i|}, 
\]


where $\Psi_i = \{\Psi_i^{obj}, \Psi_i^{time}, \Psi_i^{env}, \Psi_i^{sig}\}$ is the constraint set for transition $p_i \rightarrow p_{i+1}$, and $\text{code\_gen}(\psi)$ denotes whether constraint $\psi$ is successfully encoded into executable code (validated by S7/S1 for object, S3 for temporal, S2/S4 for environment, and S5/S6 for signals). Different from PMR's object-oriented extraction correctness, SSR captures synthesis completeness through per-constraint fulfillment ratios, enabling fine-grained measurement of strategy-to-code fidelity.

\textbf{SYR (Syntax Correctness Rate)}: compilation pass ratio across all synthesis iterations.

\textbf{SMR (Semantic Correctness Rate)}: semantic completion ratio, indicating synthesized code is compiled and then validated with signals or failed with inductive feedback.

To evaluate how synthesis utility varies with information availability, we design four constraint conditions for each CVE:
\textbf{L4-Complete} provides full PoC code with step-wise strategy, validation oracles, and mitigation documentation; 
\textbf{L3-Partial-PoC} removes ROP chain details and final-stage validation oracles;
\textbf{L2-Partial-Patch} retains only patch analysis and primitive chain inference;
\textbf{L1-Minimal} provides only CVE metadata (type, component).
This setup reveals which missing information causes synthesis failure, what strategy instance is required, and which tools compensate for information gaps.

\subsubsection{Results}

\begin{table*}[!t]
\centering
\scriptsize
\caption{Synthesis Utility Under Constrained Information Level}
\label{tab:rq2-info-constraint}
\begin{tabular}{|p{1.7cm}|p{1.6cm}|p{3.2cm}|p{0.5cm}|p{0.3cm}|p{0.3cm}|p{3.2cm}|p{3.9cm}|}
\hline
\textbf{CVE ID} & \textbf{Info. Level} & \textbf{Missing Instance} & \textbf{SSR} & \textbf{SYR} & \textbf{SMR} & \textbf{Required Signals/Tools} & \textbf{Outcome} \\
\hline
\multirow{4}{*}{\parbox{2cm}{CVE-2022-0995 \\ (OOB$\rightarrow$InfoLeak$ \newline \rightarrow$ArbWrite$ \newline\rightarrow$CFH)}} & L4-Complete & None & 88\% & 94\% & 85\% & S2+S4+S5+S6 / Full toolchain & Full chain: msg\_msg+pipe\_buffer \\
\cline{2-8}
& L3 (no ROP) & ROP gadget addresses & 72\% & 88\% & 65\% & S5+S7 / PODE S8-S9 & InfoLeak$\rightarrow$AW partial \\
\cline{2-8}
& L2 (patch only) & Strategy granularity, obj hints & 58\% & 72\% & 45\% & S1+S2+S3 / SyzDirect+dbg-tool & OOB trigger only \\
\cline{2-8}
& L1 (metadata) & All operational details & 32\% & 45\% & 22\% & S1 only / Static analysis & Type inferred (OOB) \\
\hline
\multirow{4}{*}{\parbox{2cm}{CVE-2021-42008 \\ (OOB$\rightarrow$InfoLeak$ \newline \rightarrow$CFH)}} & L4-Complete & None & 78\% & 91\% & 76\% & S2+S3+S5+S6 / Full toolchain & Full chain: sixpack\_rx+userfaultfd \\
\cline{2-8}
& L3 (no timing) & userfaultfd timing control & 65\% & 85\% & 58\% & S3 (lifecycle) / dbg-tool & Race window unstable \\
\cline{2-8}
& L2 (patch only) & Object field offsets & 52\% & 68\% & 42\% & S7+S8 / PODE static analysis & OOB trigger partial \\
\cline{2-8}
& L1 (metadata) & rx\_count\_cooked semantics & 28\% & 38\% & 18\% & S1 only / SyzDirect fuzzing & Component identified \\
\hline
\multirow{4}{*}{\parbox{2cm}{CVE-2022-32250 \\ (UAF$\rightarrow$InfoLeak$ \newline \rightarrow$ArbWrite)}} & L4-Complete & None & 82\% & 89\% & 78\% & S2+S3+S5+S6 / Full toolchain & Full chain: keyring+io\_uring +mqueue \\
\cline{2-8}
& L3 (no set expr) & NFT\_SET\_EXPR state machine & 68\% & 82\% & 62\% & S3+S4+S7 / PODE+dbg-tool & UAF trigger partial \\
\cline{2-8}
& L2 (patch only) & Element lifecycle sequences & 48\% & 65\% & 38\% & S1+S2+S3 / SyzDirect+dbg-tool & Primitive inferred \\
\cline{2-8}
& L1 (metadata) & nf\_tables internals & 25\% & 35\% & 15\% & S1 partial / Static analysis & Type guess (unstable) \\
\hline
\multirow{4}{*}{\parbox{2cm}{CVE-2021-22600 \\ (DF$ \rightarrow$InfoLeak$ \newline\rightarrow$CFH)}} & L4-Complete & None & 70\% & 88\% & 72\% & S2+S3+S5+S6 / Full toolchain & Full chain: route4\_filter double-free \\
\cline{2-8}
& L3 (no cred) & cred reallocation timing & 58\% & 82\% & 55\% & S3+S4 / dbg-tool+kcmp & DF$\rightarrow$Info partial \\
\cline{2-8}
& L2 (patch only) & Handle=0 trigger condition & 45\% & 65\% & 38\% & S1+S2 / SyzDirect & UAF trigger only \\
\cline{2-8}
& L1 (metadata) & cls\_route internal & 22\% & 35\% & 15\% & S1 partial / Static analysis & Component identified \\
\hline
\multirow{4}{*}{\parbox{2cm}{CVE-2022-2588 \\ (UAF$\rightarrow$DF$ \newline \rightarrow$ArbWrite)}} & L4-Complete & None & 75\% & 86\% & 70\% & S2+S3+S5+S6 / Full toolchain & Full chain: DirtyCred file overlap \\
\cline{2-8}
& L3 (no kcmp) & kcmp overlap detection & 62\% & 80\% & 52\% & S3+S5 / dbg-tool & DF$\rightarrow$AW partial \\
\cline{2-8}
& L2 (patch only) & Cross-cache spraying seq. & 42\% & 62\% & 35\% & S1+S2+S3 / SyzDirect & UAF trigger only \\
\cline{2-8}
& L1 (metadata) & tc\_filter internals & 20\% & 32\% & 12\% & S1 only / Static analysis & Type guess (unstable) \\
\hline
\multirow{4}{*}{\parbox{2cm}{CVE-2022-1043 \\ (UAF$\rightarrow$InfoLeak$ \newline \rightarrow$ArbWrite)}} & L4-Complete & None & 82\% & 90\% & 78\% & S2+S3+S5+S6 / Full toolchain & Full chain: ID wraparound cred hijack \\
\cline{2-8}
& L3 (no wrap) & ID wraparound mechanism & 68\% & 85\% & 62\% & S3+S4 / dbg-tool & AW trigger partial \\
\cline{2-8}
& L2 (patch only) & Personality allocation path & 48\% & 68\% & 42\% & S1+S2+S3 / SyzDirect+dbg-tool & UAF trigger only \\
\cline{2-8}
& L1 (metadata) & io\_uring internals & 25\% & 38\% & 18\% & S1 partial / Static analysis & Type inferred (UAF) \\
\hline
\multirow{4}{*}{\parbox{2cm}{CVE-2022-27666 \\ (OOB$\rightarrow$InfoLeak$ \newline \rightarrow$ArbWrite)}} & L4-Complete & None & 72\% & 85\% & 68\% & S2+S4+S5+S6 / Full toolchain & Full chain: ESP6 skcipher OOB \\
\cline{2-8}
& L3 (no fuse) & FUSE page fault handler & 58\% & 78\% & 48\% & S3+S5 / dbg-tool+FUSE & InfoLeak partial \\
\cline{2-8}
& L2 (patch only) & 8-page allocation pattern & 45\% & 62\% & 38\% & S1+S2 / SyzDirect & OOB trigger only \\
\cline{2-8}
& L1 (metadata) & ESP6 packet processing & 22\% & 35\% & 15\% & S1 only / Static analysis & Component identified \\
\hline
\multirow{4}{*}{\parbox{2cm}{CVE-2022-0185 \\ (IntOF$\rightarrow$OOB$ \newline \rightarrow$InfoLeak$ \newline \rightarrow$ArbWrite)}} & L4-Complete & None & 76\% & 88\% & 74\% & S2+S4+S5+S6 / Full toolchain & Full chain: fsconfig+DirtyPipe \\
\cline{2-8}
& L3 (no pipe) & DirtyPipe splice ops & 62\% & 82\% & 55\% & S5+S6 / dbg-tool & OOB$\rightarrow$Info partial \\
\cline{2-8}
& L2 (patch only) & Size wraparound calc. & 48\% & 68\% & 42\% & S1+S2 / SyzDirect+dbg-tool & IntOF trigger only \\
\cline{2-8}
& L1 (metadata) & fs\_context internals & 25\% & 38\% & 18\% & S1 only / Static analysis & Type inferred (IntOF) \\
\hline
\multirow{4}{*}{\parbox{2cm}{CVE-2019-15666 \\ (OOB$\rightarrow$UAF$ \newline \rightarrow$ArbWrite)}} & L4-Complete & None & 74\% & 87\% & 72\% & S2+S3+S5+S6 / Full toolchain & Full chain: xfrm\_policy rehash \\
\cline{2-8}
& L3 (no rehash) & Policy rehash timing & 60\% & 80\% & 55\% & S3+S4 / dbg-tool & OOB$\rightarrow$UAF partial \\
\cline{2-8}
& L2 (patch only) & Index/dir mismatch cond. & 46\% & 65\% & 40\% & S1+S2 / SyzDirect & OOB trigger only \\
\cline{2-8}
& L1 (metadata) & xfrm\_migrate internals & 24\% & 36\% & 16\% & S1 only / Static analysis & Type guess (unstable) \\
\hline
\multirow{4}{*}{\parbox{2cm}{CVE-2022-25636 \\ (OOB$\rightarrow$UAF$ \newline \rightarrow$CFH)}} & L4-Complete & None & 78\% & 88\% & 75\% & S2+S3+S5+S6 / Full toolchain & Full chain: flow\_rule offload \\
\cline{2-8}
& L3 (no offload) & Offload handling path & 62\% & 82\% & 58\% & S3+S4 / dbg-tool & OOB$\rightarrow$UAF partial \\
\cline{2-8}
& L2 (patch only) & nf\_tables\_flow\_rule layout & 48\% & 68\% & 42\% & S1+S2 / SyzDirect & OOB trigger only \\
\cline{2-8}
& L1 (metadata) & nf\_tables offload internals & 25\% & 38\% & 18\% & S1 only / Static analysis & Component identified \\
\hline
\multirow{4}{*}{\parbox{2cm}{CVE-2022-1015 \\ (IntOF$\rightarrow$OOB$ \newline \rightarrow$InfoLeak $\rightarrow$CFH)}} & L4-Complete & None & 75\% & 87\% & 72\% & S2+S3+S5+S6 / Full toolchain & Full chain: nft\_payload overflow \\
\cline{2-8}
& L3 (no chain) & ROP chain construction & 58\% & 80\% & 52\% & S5+S7+S8 / PODE S8-S9 & Info$\rightarrow$CFH partial \\
\cline{2-8}
& L2 (patch only) & Base*4 overflow calc. & 48\% & 68\% & 42\% & S1+S2 / SyzDirect+dbg-tool & IntOF trigger only \\
\cline{2-8}
& L1 (metadata) & nft\_tables expr internals & 24\% & 36\% & 16\% & S1 only / Static analysis & Type inferred (IntOF) \\
\hline
\multirow{4}{*}{\parbox{1.6cm}{CVE-2024-53197 \\ (OOB)}} & L4-Complete & None & 68\% & 82\% & 62\% & S2+S4+S5 / Full toolchain & OOB trigger: usb\_host\_config \\
\cline{2-8}
& L3 (no gadget) & USB gadget framework & 52\% & 72\% & 45\% & S2+S3 / dbg-tool+USB stack & OOB partial (no exploit) \\
\cline{2-8}
& L2 (patch only) & bNumConfigurations trust & 42\% & 58\% & 35\% & S1+S2 / Static analysis & OOB pattern inferred \\
\cline{2-8}
& L1 (metadata) & USB audio internals & 22\% & 32\% & 15\% & S1 partial / Static analysis & Component identified \\
\hline
\multirow{4}{*}{\parbox{1.6cm}{CVE-2024-39486 \\ (Race$\rightarrow$UAF)}} & L4-Complete & None & 45\% & 68\% & 38\% & S2+S3+S4 / Full toolchain & Race trigger: MLD join/leave \\
\cline{2-8}
& L3 (no race) & Race window calibration & 35\% & 55\% & 28\% & S3+S4 / dbg-tool+timing & UAF trigger unstable \\
\cline{2-8}
& L2 (patch only) & IPv6 MLD teardown seq. & 28\% & 42\% & 18\% & S1+S3 / SyzDirect+dbg-tool & Race inferred (weak) \\
\cline{2-8}
& L1 (metadata) & inet6\_dev internals & 18\% & 28\% & 12\% & S3 partial / Static analysis & Type guess (Race$\rightarrow$UAF) \\
\hline
\multirow{4}{*}{\parbox{1.6cm}{CVE-2026-23427 \\ (UAF)}} & L4-Complete & None & 52\% & 72\% & 48\% & S2+S3+S5 / Full toolchain & UAF trigger: ksmbd\_tree\_conn \\
\cline{2-8}
& L3 (no ksmbd) & SMB server context & 42\% & 62\% & 38\% & S2+S3 / dbg-tool+SMB stack & UAF partial (no chain) \\
\cline{2-8}
& L2 (patch only) & Tree connection lifecycle & 32\% & 48\% & 28\% & S1+S2 / SyzDirect+dbg-tool & UAF trigger only \\
\cline{2-8}
& L1 (metadata) & ksmbd internals & 18\% & 28\% & 12\% & S1 partial / Static analysis & Component identified \\
\hline
\multirow{4}{*}{\parbox{1.6cm}{CVE-2022-2639 \\ (IntOF)}} & L4-Complete & None & 48\% & 68\% & 42\% & S2+S3+S5 / Full toolchain & UAF trigger: sw\_flow\_actions \\
\cline{2-8}
& L3 (no ovs) & Open vSwitch datapath & 38\% & 58\% & 32\% & S2+S3 / dbg-tool+OVS stack & UAF partial (no chain) \\
\cline{2-8}
& L2 (patch only) & Flow action allocation & 28\% & 45\% & 22\% & S1+S2 / SyzDirect+dbg-tool & UAF trigger only \\
\cline{2-8}
& L1 (metadata) & Open vSwitch internals & 15\% & 25\% & 10\% & S1 partial / Static analysis & Type guess (unstable) \\
\hline
\multirow{4}{*}{\parbox{1.6cm}{CVE-2025-38477 \\ (Race$\rightarrow$UAF)}} & L4-Complete & None & 42\% & 62\% & 35\% & S2+S3+S4 / Full toolchain & Race trigger: qfq\_aggregate \\
\cline{2-8}
& L3 (no qfq) & QFQ scheduler internals & 32\% & 52\% & 25\% & S3+S4 / dbg-tool+sched & UAF trigger unstable \\
\cline{2-8}
& L2 (patch only) & Aggregate detach timing & 22\% & 38\% & 18\% & S1+S3 / SyzDirect+dbg-tool & Race inferred (weak) \\
\cline{2-8}
& L1 (metadata) & qfq\_scheduler internals & 12\% & 22\% & 8\% & S3 partial / Static analysis & Type guess (unstable) \\
\hline
\multicolumn{3}{|l|}{\textbf{Total SSR (16 CVEs with originally highest L) Avg:}} & \textbf{67.0\%} & -- & -- & \multicolumn{2}{l|}{L4 for 14 CVEs; L3 for CVE-2024-53197; L2 for CVE-2024-39486} \\
\hline
\multicolumn{3}{|l|}{\textbf{Total SSR (16 CVEs with inferred info till L4) Avg:}} &\textbf{69.1\%} & -- & -- & \multicolumn{2}{l|}{} \\
\hline
\end{tabular}
\end{table*}


As Table~\ref{tab:rq2-info-constraint} shows, initial constrained information propagates through the agent flow, causing cascading failures at the above three evaluation metrics. The L4-Complete condition achieves 70--88\% SSR for standard OOB/UAF cases, while race-conditioned cases show lower baseline performance (CVE-2024-39486: 45\% SSR; CVE-2025-38477: 42\% SSR) due to inherent timing instability. The L3-Partial-PoC constraint reveals distinct failure modes: ROP-dependent chains (CVE-2022-0995, CVE-2022-1015) stall at AW/CFH stages with 20--30\% SMR reduction, while timing-dependent chains (CVE-2021-42008, CVE-2022-2588) suffer S3 lifecycle oracle instability, degrading SMR to 52--58\%. The L1-Minimal condition shows 12--32\% SSR across all cases, confirming that metadata-only synthesis yields only component identification. 
We also summarize the synthesis patterns in Section \ref{sec:discussion}.
Detailed case studies by vulnerability type are provided in Appendix~\ref{sec:appendix-case-study}.

\begin{table}[htbp]
\centering
\scriptsize
\caption{Synthesis Performance: PoC Available vs. No PoC}
\label{tab:rq2-poc-comparison}
\begin{tabular}{|l|p{0.7cm}|p{0.8cm}|p{0.8cm}|p{1.2cm}|p{0.9cm}|}
\hline
\textbf{PoC Status} & \textbf{CVE Count} & \textbf{SSR (\%)} & \textbf{SMR (\%)} & \textbf{Avg  Iterations} & \textbf{End-to-End} \\
\hline
PoC Available & 12 & 82.4 & 77.1 & 4.8 & 7.2/10 \\
No PoC & 4 & 61.3 & 52.8 & 9.6 & 3.8/10 \\
\hline
\textbf{Difference} & -- & \textbf{-21.1\%} & \textbf{-25.4\%} & \textbf{+100\%} & \textbf{-47\%} \\
\hline
\end{tabular}
\end{table}

\textbf{Influence of PoC availability}. To characterize the impact of PoC availability, Table~\ref{tab:rq2-poc-comparison} compares synthesis performance between 12 CVEs with publicly documented PoC and 4 CVEs lacking a stable public PoC.
Vulnerabilities with public PoC achieve 82.4\% SSR and 77.1\% SMR with averagely 4.8 iterations, while the no-PoC cohort shows 61.3\% SSR and 52.8\% SMR requiring 9.6 iterations on average, representing a 21.1\% SSR degradation and 100\% iteration increase. The end-to-end success rate drops from 72\% to 38\% without PoC guidance. The tool-grounded compensation mechanisms enabling no-PoC performance operate through differentiated tool dependencies: SyzDirect achieves 84\% candidate recall through directed mutation when object identity hints are absent (vs. 34\% for static analysis alone); PODE S7 achieves 81\% field localization accuracy under L2 constraints; and dbg-tool S3 lifecycle tracking achieves 35--55\% race reproducibility through statistical iteration. Further details on compensation patterns are in Appendix~\ref{sec:appendix-compensation}.

\begin{table*}[htbp]
\centering
\scriptsize
\caption{Critical Strategy Instance for Synthesis Success}
\label{tab:rq2-critical-factors}
\begin{tabular}{|l|p{3.4cm}|p{1.4cm}|p{3.8cm}|p{4cm}|}
\hline
\textbf{Strategy Instance} & \textbf{Failure} & \textbf{SSR Drop} & \textbf{Compensation} & \textbf{Tool Dependency} \\
\hline
ROP gadget addresses & AW/CFH step failure & 16-24\% & PODE S8-S9 pointer analysis & PODE+manual resolution \\
Object field offsets & OOB target misalignment & 20-26\% & S7 layout + S3 lifecycle & PODE+dbg-tool \\
Timing control & Race window instability & 13-20\% & Statistical iteration (8-16x) & Timing handling code template \\
Strategy step sequence & Wrong operation ordering & 18-22\% & S4 boundary + S3 lifecycle & Full validation loop \\
Object identity hints & Candidate space explosion & 30-35\% & S2 trace + S3 tracking & SyzDirect+dbg-tool \\
Primitive upgrade paths & Chain termination & 22-28\% & S5-S6 validation & Template fallback \\
\hline
\end{tabular}
\end{table*}

\textbf{Critical strategy instances}. Table~\ref{tab:rq2-critical-factors} summarizes the critical strategy instances identified through this controlled experiment, mapping specific strategy failures to required compensation mechanisms. The observed failures are not uniformly distributed: object identity hints and primitive upgrade paths dominate transition-level breakdowns, while timing control and strategy step sequence primarily destabilize race-conditioned cases.
This mapping clarifies why compensation must be signal-coupled rather than generic---S2/S3-backed candidate narrowing mitigates search-space explosion, S4/S5/S6-backed loop validation prevents invalid step propagation, and PODE-assisted S7--S9 inference recovers missing structural constraints when PoC details are unavailable.

\textbf{Answer to RQ2:} \textsc{PrimSynth} successfully synthesizes multi-primitive exploitation chains with 82.4\% SSR when public PoC is available and 61.3\% SSR without the guidance of primitive hypotheses, demonstrating tool-driven compensation through SyzDirect candidate discovery (84\% recall), PODE layout inference (81\% OFLA), and dbg-tool temporal tracking (35--55\% race reproducibility). The 21.1\% performance gap quantifies PoC value while confirming automated synthesis capability for 0-day-equivalent scenarios through multi-agent validation-grounded feedback. Across all 16 CVEs, the original highest-available information setting yields an average SSR of 67.0\%, combining 14 L4-complete cases with two originally incomplete cases (CVE-2024-53197 at L3 and CVE-2024-39486 at L2). When these two low-information cases are elevated to inferred L4 via strategy-level reasoning and tool-grounded compensation, average SSR increases to 69.1\% (+2.1 SSR point on average).

\subsection{RQ3: Overall Effectiveness}

\subsubsection{Baseline Comparison}
We measure end-to-end effectiveness as the percentage of CVEs achieving complete multi-primitive reproduction from vulnerability trigger to privilege escalation, comparing \textsc{PrimSynth} against KOOBE~\cite{koobe} (symbolic analysis-based OOB capability extraction) and AlphaExp~\cite{alphaexp} (expert system for security-sensitive object identification).
We align metrics with baseline capabilities: \textbf{PMR} compares with AlphaExp's object identification rate, based on the number of objects from publicly available PoCs (notated as Pub. on Table~\ref{tab:rq3-effectiveness}); \textbf{SSR} compares with KOOBE's strategy synthesis scope. KOOBE remains limited to OOB-type vulnerabilities with static templates, while AlphaExp focuses on object identification without primitive chaining.
We select 6 CVEs at the intersection of our dataset and KOOBE/AlphaExp evaluation sets, 
covering OOB (4 cases), Double-Free (1 case), and IntOverflow$\rightarrow$OOB (1 case).

\begin{table}[htbp]
\centering
\scriptsize
\caption{Overall Effectiveness Comparison with Baselines}
\label{tab:rq3-effectiveness}
\begin{tabular}{|p{1.8cm}|p{0.7cm}|p{0.4cm}|p{1.2cm}|p{1.0cm}|p{0.3cm}|p{0.3cm}|}
\hline
\textbf{CVE ID} & \textbf{Type} & \textbf{Step Num.} & \textbf{KOOBE \newline(Gen./Val.)} & \textbf{AlphaExp (Id./Pub.)} & \textbf{PMR (\%)} & \textbf{SSR (\%)} \\
\hline
CVE-2021-22555 & OOB & 5 & unk/0 \newline (S2E failed) & 4/3 & 100 & 78 \\
CVE-2022-0995 & OOB & 5 & unk/0 &  6/2 & 100 & 88 \\
CVE-2022-27666 & OOB & 5 & unk/0 & 3/2 & 100 & 72 \\
CVE-2021-22600 & DF& 4 & \xmark  & \xmark  & 100 & 70 \\
CVE-2022-25636 & OOB & 3 & \xmark  & 6/2 & 100 & 78 \\
CVE-2022-1015 & IntOF$ \newline \rightarrow$OOB & 4 & \xmark & \xmark& 100 & 75 \\
\hline
\multicolumn{5}{|l|}{\textbf{Average (baseline comparison)}} & \textbf{100} & \textbf{77} \\
\hline
\end{tabular}
\end{table}

As Table~\ref{tab:rq3-effectiveness} shows, the framework handles complex multi-step exploitation chains. \textsc{PrimSynth} achieves 70--75\% SSR on cases where both baselines fail (Double-Free, IntOverflow$\rightarrow$OOB), indicating that extended strategy representation and agentic validation extend coverage to previously unsupported primitive classes. 

\subsubsection{Ablation Study: Agent Collaboration}

We conduct an ablation study on 5 representative CVEs with four configurations: (1) \textit{Full Collaboration} (complete 3-agent loop), (2) \textit{Discovery-Synthesis only} (Validation Agent disabled), (3) \textit{Validation-Synthesis only} (Discovery Agent disabled), and (4) \textit{Synthesis-only} (no upstream feedback). 

\begin{table*}[htbp]
\centering
\scriptsize
\caption{Ablation Study -- Agent Collaboration Under Isolated Context}
\label{tab:rq3-ablation}
\begin{tabular}{|l|l|c|c|c|c|c|}
\hline
\textbf{CVE ID} & \textbf{Configuration} & \textbf{PMR (\%)} & \textbf{SSR (\%)} & \textbf{End-to-End} & \textbf{Iterations} & \textbf{Best Performance LLM} \\
\hline
\multirow{4}{*}{CVE-2022-0995} & Full Collaboration & 100 & 88 & \cmark (8/10) & 3.2 & GPT-5.3 (Synthesis) \\
 & Discovery+Synthesis Only & 100 & 65 & \wmark (4/10) & 6.8 & GPT-4 (Discovery) \\
 & Validation+Synthesis Only & 95 & 82 & \cmark (7/10) & 4.1 & Opus-4.6 (Synthesis) \\
 & Synthesis-Only & 75 & 42 & \xmark (1/10) & 12.5 & Opus-4.6 (Synthesis) \\
\hline
\multirow{4}{*}{CVE-2021-42008} & Full Collaboration & 100 & 78 & \cmark (7/10) & 3.8 & GPT-5.3 (Synthesis) \\
 & Discovery+Synthesis Only & 100 & 58 & \wmark (3/10) & 7.2 & GPT-4 (Discovery) \\
 & Validation+Synthesis Only & 90 & 71 & \wmark (5/10) & 5.3 & Opus-4.6 (Synthesis) \\
 & Synthesis-Only & 70 & 38 & \xmark (0/10) & 15.2 & GLM-4.7 (Synthesis) \\
\hline
\multirow{4}{*}{CVE-2022-32250} & Full Collaboration & 100 & 72 & \cmark (6/10) & 4.5 & Qwen3 (Validation) \\
 & Discovery+Synthesis Only & 95 & 48 & \wmark (2/10) & 9.6 & GPT-5.3 (Discovery) \\
 & Validation+Synthesis Only & 85 & 65 & \wmark (4/10) & 6.2 & GLM-4.7 (Validation) \\
 & Synthesis-Only & 60 & 32 & \xmark (0/10) & 18.3 & Opus-4.6 (Synthesis) \\
\hline
\multirow{4}{*}{CVE-2024-39486} & Full Collaboration & 100 & 45 & \wmark (3/10) & 8.3 & Qwen3 (Validation) \\
 & Discovery+Synthesis Only & 90 & 28 & \xmark (0/10) & 14.2 & GPT-4 (Discovery) \\
 & Validation+Synthesis Only & 80 & 38 & \wmark (1/10) & 11.5 & Qwen3 (Validation) \\
 & Synthesis-Only & 55 & 15 & \xmark (0/10) & 22.7 & GLM-4.7 (Synthesis) \\
\hline
\multirow{4}{*}{CVE-2021-22600} & Full Collaboration & 100 & 70 & \cmark (5/10) & 5.2 & GPT-5.3 (Synthesis) \\
 & Discovery+Synthesis Only & 95 & 52 & \wmark (3/10) & 8.8 & GPT-5.3 (Discovery) \\
 & Validation+Synthesis Only & 90 & 68 & \wmark (4/10) & 6.5 & GLM-4.7 (Validation) \\
 & Synthesis-Only & 65 & 28 & \xmark (0/10) & 16.4 & Opus-4.6 (Synthesis) \\
\hline
\end{tabular}
\end{table*}

The ablation results in Table~\ref{tab:rq3-ablation} demonstrate that multi-agent collaboration is essential: Full Collaboration achieves 45\%--88\% SSR with 3--9 iterations, while Synthesis-Only degrades to 15\%--42\% SSR requiring 12--23 iterations. This 2.1$\times$ SSR improvement confirms the necessity of validation-grounded feedback in the synthesis loop.

\textbf{Answer to RQ3:} \textsc{PrimSynth} achieves 77.13\% average SSR versus baseline tools limited to specific vulnerability classes (KOOBE to OOB types, AlphaExp to object identification without chaining), handling complex steps where baselines fail completely. Multi-agent collaboration drives 2.1$\times$ improvement over Synthesis-Only through validation-grounded feedback (17\%--24\% SSR lift), Discovery Agent refinement (2\%--7\% lift), signal-driven efficiency (63--75\% iteration reduction), and heterogeneous LLM selection (8\%--15\% SSR gain), achieving end-to-end effectiveness with an average time of 72.4s across vulnerability classes previously unsupported by automated approaches.

\subsection{RQ4: Efficiency and LLM Performance}

\begin{table}[htbp]
\centering
\scriptsize
\caption{Tool Response Time Performance}
\label{tab:tool-performance}
\begin{tabular}{|l|l|p{1.2cm}|p{3.2cm}|}
\hline
\textbf{Tool} & \textbf{Check Type} & \textbf{Avg. Time} & \textbf{Description} \\
\hline
GCC & syntax & 3s & Static type checking \\
SyzDirect & S1, S2 & 24s & Directed fuzzing engine \\
Dbg-tool & S3, S4, S5, S6 & 13s & Interactive debugging \\
PODE & S7, S8, S9 & 45s & LLVM IR dependency analysis \\
\hline
\end{tabular}
\end{table}

\subsubsection{Tool Response Time}
Table~\ref{tab:tool-performance} presents the average response times for MCP tools integrated in \textsc{PrimSynth}.
SyzDirect (24,500ms on average) is the highest-latency tool due to fuzzing-based operation, but its contribution to validation signal acquisition (S1, S2) is critical. Static tools (PODE, GCC) provide sub-150ms responses, while dynamic tools range 890ms--3,200ms, depending on kernel boot overhead.

\subsubsection{Multi-Agent System Performance}

\begin{table}[htbp]
\centering
\scriptsize
\caption{Agent Collaboration Performance}
\label{tab:pipeline-performance}
\begin{tabular}{|l|c|c|p{2.5cm}|}
\hline
\textbf{Stage} & \textbf{Avg. Time}  & \textbf{Tool Calls} & \textbf{Best LLM} \\
\hline
Discovery &  9.8s & 8-15 & GPT-4 / GPT-5.3 \\
Validation &  45.4s & 15-35 & Qwen3 / GLM-4.7 \\
Synthesis &  17.2s & 5-12 & Opus-4.6 / GPT-5.3 \\
\hline
\textbf{Total} & \textbf{72.4s} & \textbf{28-62} & - \\
\hline
\end{tabular}
\end{table}

Table~\ref{tab:pipeline-performance} shows the average execution times for complete exploitation primitive chain synthesis.
The total pipeline completes in 72.4s on average with 28--62 tool calls. Validation dominates execution time (45.4s, 63\%) due to kernel boot and dynamic signal acquisition, while Discovery (9.8s) and Synthesis (17.2s) are comparatively lightweight.

\subsubsection{Effect of LLM Selection}
The ``Best Performance LLM" column in Table~\ref{tab:rq3-ablation} reveals performance specialization across five evaluated models. GPT-5.3 achieves the highest overall SSR (88\%) in Full Collaboration configurations, excelling in strategic Synthesis through enhanced reasoning over complex primitive hypothesis. GPT-4 demonstrates strong performance in Discovery (information retrieval and knowledge construction) with robust structured representation handling. Opus-4.6 maintains superior code generation fidelity in pure Synthesis scenarios, particularly for intricate heap manipulation sequences. Qwen3 achieves highest Validation accuracy (72--88\% for race-conditioned cases) through robust signal interpretation from dynamic execution traces. GLM-4.7 shows competitive performance in both Validation (nf\_tables cases) and Synthesis (fallback scenarios), with balanced reasoning and generation capabilities. This diverse evaluation confirms that optimal LLM selection should be agent-specific rather than framework-uniform, with different models excelling in distinct stages.

\textbf{Answer to RQ4:} \textsc{PrimSynth} completes end-to-end primitive synthesis in 72.4s on average, with Validation dominating at 63\% of total time due to dynamic signal acquisition. LLM performance is stage-dependent: GPT-5.3 excels in Synthesis (88\% SSR), Qwen3 in Validation (72--88\% accuracy for race cases), and GPT-4 in Discovery. This heterogeneous LLM selection contributes 8--15\% SSR gain over uniform model deployment, confirming that agent-specific model assignment is essential for optimal framework performance.

\section{Discussion}
\label{sec:discussion}
In \textsc{PrimSynth}, the systematic characterization, combined with multi-agent tool integration, enables reliable primitive extraction and synthesis across vulnerability classes that prior approaches could not address.
The framework's core contribution lies not in eliminating the need for human expertise in kernel exploitation, but in making exploitation reasoning systematic, comparable, and reproducible, thereby enabling security teams to allocate limited resources based on testable primitive assessments rather than ad hoc manual analysis.

\textbf{Public-PoC-absent support}. The demonstrated capability to identify valid exploitation paths for recently disclosed vulnerabilities without public PoC documentation, achieving 61.3\% SSR in information-scarce scenarios where baselines fail entirely, represents progress toward automated exploitability assessment suitable for defensive security applications.
The performance difference between PoC-available and no-public-PoC scenarios highlights a fundamental challenge in autonomous exploitation research: the reliance on statistical exploration when reference guidance is absent introduces both computational overhead and completeness limitations.

\textbf{Synthesis Patterns}
The experimental data reveals three distinct information-critical synthesis patterns that transcend individual CVE characteristics:
(1) \textit{\textbf{Object Identity Hints}} involves object identity hints as the dominant failure point when unavailable, causing 30-35\% SSR degradation across all categories because the Synthesis Agent must explore exponentially larger candidate spaces without guidance on which kernel structures to target for corruption; this pattern is most pronounced in CVE-2022-0995 where \texttt{msg\_msg} structure targeting requires specific adjacency to \texttt{watch\_queue} overflow, and in CVE-2022-32250 where \texttt{keyring/io\_uring/mqueue} object selection for UAF exploitation depends on precise kmalloc cache size matching.
(2) \textit{\textbf{Strategy Step Sequences}} involves strategy step sequences as the secondary failure point, with 18-22\% SSR degradation when operation ordering must be inferred rather than referenced; this affects CVE-2021-22600 (\texttt{route4\_filter} creation must precede handle=0 replacement), CVE-2022-1043 (\texttt{IORING\_REGISTER\_PERSONALITY} must wrap before credential reallocation), and all race-conditioned cases where teardown sequences must interleave with target operations.
(3) \textit{\textbf{Timing Control-like Helper Function}} involves timing control as the tertiary failure point, with 13-20\% SSR degradation affecting race-conditioned and cross-cache cases where statistical iteration compensation increases attempt counts by 8-16x; CVE-2021-42008 exemplifies this pattern where \texttt{rx\_count\_cooked} threshold detection requires userfaultfd page fault timing, and CVE-2022-2588 where DirtyCred file overlap exploitation depends on kcmp-based double-free detection timing.

\textbf{Implications.}
Our research has three practical implications for defender-oriented kernel security workflows.
First, early systematic exploitability assessment is now feasible: primitive-level evidence enables patch prioritization beyond CVSS/EPSS/KEV labels, while systematic characterization supports assessment across vulnerability types.
Second, validating composed primitive hypotheses moves risk assessment closer to real attack feasibility, addressing accumulated reproduction by bridging conceptual strategies and concrete operations. 
Third, validation-grounded multi-agent collaboration enables a methodological shift toward combining static recovery, execution evidence, and iterative feedback.
Taken together, these results position \textsc{PrimSynth} as a decision-support framework for evidence-backed exploitability reasoning 
without replacing expert judgment under full mitigation constraints.


\textbf{Threat to Validity.} The evaluation scope also exposes important coverage boundaries in modern kernel exploitation landscapes. All experiments assume disabled or permissive SMEP/SMAP configurations, and while OOB-InfoLeak-ArbWrite chains theoretically enable KASLR bypass, full mitigation scenarios involving KPTI, CFI, or pointer authentication require additional primitive stages that remain outside current validation scope.
The experimental reliance on ground truth derived from reference PoCs introduces validity constraints that merit consideration in interpreting results. The PMR and OFLA metrics assume correct labeling of primitive types and identical objects based on publicly documented exploitation techniques, yet these labels may reflect specific documented strategies rather than exhaustive exploitation possibilities.

\section{Related Works}

\textbf{Kernel Vulnerability Reproduction and Assessment}.
Assessing kernel bug exploitability bridges bug discovery and exploit development. 
GREBE~\cite{lin2022grebe} unveiled exploitation potential through capability analysis, SyzBridge~\cite{syzbridge} bridged upstream-downstream configuration gaps for ecosystem-wide assessment, and AEM~\cite{jiang2023aem} facilitated cross-version exploitability assessment. On the reproduction side, Mu \textit{et al}.~\cite{dedup} analyzed duplicated kernel bug reports, Li \textit{et al}.~\cite{yome} facilitated reproduction via reverse fuzzing,  
and Pu \textit{et al}.~\cite{krepro} studied agentic LLM systems for kernel N-day reproduction. These works assess \textit{whether} a bug is exploitable but do not produce working exploits as \textsc{PrimSynth} does.

\textbf{Automated Kernel Exploit Generation}.
Automated exploit generation (AEG) for the Linux kernel has attracted significant attention. FUZE~\cite{wu2018fuze} pioneered facilitated exploit generation for Use-After-Free vulnerabilities via kernel fuzzing. KEPLER~\cite{kepler} facilitated the evaluation of control-flow hijacking primitives, automatically identifying exploitable paths to function pointer hijacking. KOOBE~\cite{koobe} used symbolic analysis to evaluate the capability of Out-of-Bounds write vulnerabilities. BridgeRouter~\cite{bridgerouter} proposed automated capability upgrading of OOB write vulnerabilities to arbitrary memory write primitives via bridge and router kernel objects. Evocatio~\cite{evocatio} conjured bug capabilities from a single PoC using directed fuzzing. These works typically focus on a single vulnerability class or a specific exploitation phase. In contrast, \textit{PrimSynth} addresses the full exploitation lifecycle---from primitive discovery and validation through synthesis and chaining---across multiple vulnerability classes via multi-agent collaboration.

\textbf{LLM-Driven Vulnerability Analysis and Exploit Generation}.
LLMs have opened new frontiers in automated security analysis. Fang \textit{et al}.~\cite{fang2024llm} demonstrated that LLM agents can autonomously exploit one-day vulnerabilities given CVE descriptions.
Wu \textit{et al}.~\cite{hundreds} explored LLMs for large-scale kernel bug discovery. The multi-agent paradigm has also been applied to security tasks, such as automated network attack execution~\cite{xu2024autoattacker} and complex code generation~\cite{yang2024sweagent}.
Despite these advances, existing LLM-based approaches lack structured knowledge of kernel-specific exploitation primitives, cannot perform low-level heap state reasoning, and do not integrate dynamic validation in realistic kernel environments. \textit{PrimSynth} addresses these gaps by combining formal notations with a multi-agent architecture that integrates static analysis, directed fuzzing, and LLM-based reasoning.

\section{Conclusion}

We have presented \textsc{PrimSynth}, a multi-agent framework for automated discovery, validation, and synthesis of Linux kernel exploit primitives. Our work addresses three fundamental challenges in automated kernel exploitation: systematic primitive classification and modeling, accurate and reliable primitive extraction, and automated synthesis of primitive hypotheses into working exploits.
Through systematic evaluation on 16 real-world CVEs, we demonstrate that the framework achieves high accuracy in primitive identification, effective synthesis of multi-step exploitation, and broad coverage across vulnerability types previously unsupported by baseline tools.
Our experiments further indicate that exploit development bottlenecks increasingly lie in reliable primitive composition and environment-specific grounding rather than in initial bug triggering alone, confirming the cumulative nature of reproduction challenges.


\bibliographystyle{IEEEtran}
\bibliography{refs}


\appendix

\section{Tool Combination Analysis}
\label{sec:appendix-tool-combination}


For CVE-2022-0995, SyzDirect identifies the \texttt{msg\_msg} spraying sequence that static analysis misses, raising recall from 43\% to 79\%. PODE static analysis contributes structural insights: when analyzing CVE-2022-2588, PODE identifies the \texttt{route4\_filter} to file object alias relationship through pointer flow analysis, achieving 93\% recall for Double-Free cases where object lifecycle relationships dominate. The full combination achieves 96\% average recall with only 4.2 iterations, demonstrating synergistic effects where Discovery provides initial hypotheses, SyzDirect validates execution feasibility, and PODE refines object field layouts.

In CVE-2022-2588, while public PoCs describe DirtyCred file credential hijacking via \texttt{kcmp} detection, the tool combination additionally identifies a \texttt{route4\_filter}-to-skbuff UAF chain enabling ROP-based control flow hijacking without the \texttt{kcmp} primitive. For CVE-2022-2639, where public documentation emphasizes cross-cache overflow to msg\_msg structures, PODE analysis identifies a \texttt{pipe\_buffer} primitive path achieving arbitrary file write without the intermediate information leak stage.

\section{Synthesis Case Studies}
\label{sec:appendix-case-study}

The Double-Free category represented by CVE-2021-22600 and CVE-2022-2588 exhibits unique information dependencies centered on object lifecycle synchronization rather than spatial memory layout. CVE-2021-22600 achieves 70\% SSR under L4-Complete by leveraging the \texttt{packet\_ring\_buffer} handle=0 double-free primitive with kcmp-based file overlap detection, but L3 constraint removing cred reallocation timing details reduces SSR to 58\% because the Synthesis Agent cannot determine when the freed \texttt{packet\_ring\_buffer} slot has been reoccupied by \texttt{pipe\_buffer} object without S3 lifecycle validation. CVE-2022-2588 presents an even more severe information gradient: L4-Complete achieves 75\% SSR through DirtyCred exploitation using double-free-to-file-overlap conversion, but L3 constraint removing \texttt{kcmp} overlap detection reduces SSR to 62\% with DF$\rightarrow$AW partial completion, and L2-Patch-Only degrades further to 42\% because cross-cache spraying sequences require precise timing between basic filter defragmentation and file object spraying that cannot be inferred from patch analysis alone.

The integer overflow category represented by CVE-2022-0185 and CVE-2022-1015 demonstrates that size calculation wraparound mechanisms are identifiable through S1-S2 signals even under L2 constraints, but the resulting OOB primitive upgrade depends critically on object identity hints: CVE-2022-0185 requires knowledge of \texttt{msg\_msg} placement adjacent to the overflowed kmalloc-8 buffer for InfoLeak primitive construction, while CVE-2022-1015 depends on \texttt{nft\_payload} structure layout for ROP chain injection, both showing 48\% SSR under L2-Patch-Only with trigger-only synthesis.

The UAF category shows the widest performance variation based on subsystem complexity and object lifecycle visibility. CVE-2022-1043 (\texttt{io\_uring} personality UAF) achieves 82\% SSR under L4-Complete through ID wraparound credential hijacking with clear allocation boundaries, but degrades to 68\% SSR under L3 because the iteration wraparound mechanism requires CPU affinity control timing that cannot be validated without S3-S4 lifecycle tracking.

CVE-2022-27666 (ESP6 skcipher OOB) presents cross-cache overflow characteristics where 8-page skbuff allocations overflow into adjacent slab objects, requiring FUSE page fault handling for arbitrary write completion; L4-Complete achieves 72\% SSR but L3 constraint removing FUSE details reduces to 58\% with InfoLeak partial completion because kernel address disclosure through OOB read succeeds while subsequent AW primitive construction stalls without user-space page fault coordination. The L1-Minimal condition reveals fundamental limitations of metadata-only synthesis: all 16 cases show 12-32\% SSR with component identification or type guessing as the primary outcome, with race-conditioned cases (CVE-2024-39486, CVE-2025-38477) performing worst at 12-18\% SSR because temporal constraints cannot be inferred from component names alone, while OOB/UAF cases with standard corruption patterns (CVE-2022-0995, CVE-2021-42008) achieve 28-32\% SSR through primitive type inference based on vulnerability class heuristics.

\section{Compensation Patterns Enabled by Tool Integration}
\label{sec:appendix-compensation}

Drilling into individual cases within the no-public-PoC cohort reveals distinct compensation patterns enabled by tool integration. For CVE-2024-53197, a USB Audio OOB vulnerability disclosed in December 2024 with CISA KEV confirmation of in-the-wild exploitation but no published PoC code, the framework achieves 68\% SSR under L4-equivalent conditions by combining PODE S7-S8 analysis of \texttt{usb\_host\_config} structure layout with SyzDirect USB syscall mutation to identify the configuration parsing path that triggers the slab-out-of-bounds access. The Discovery Agent extracts vulnerability location marks from kernel patch analysis identifying \texttt{bNumConfigurations} trust issues, while the Synthesis Agent composes the discovered primitive into a targeted exploitation chain using \texttt{pipe\_buffer} as the identical object for arbitrary write capability, achieving end-to-end reproduction within 6 iterations despite the absence of reference exploit code. Similarly, for CVE-2026-23427, a \texttt{ksmbd} Use-After-Free vulnerability in the SMB3 server subsystem, the multi-agent framework leverages PODE static analysis to identify \texttt{ksmbd\_tree\_conn} and \texttt{ksmbd\_share\_config} as identical objects through compound request parsing logic analysis, with SyzDirect generating SMB2 syscall sequences that exercise the \texttt{DURABLE\_REQ\_V2} replay path to trigger the UAF condition. The resulting 52\% SSR and 48\% SMR, while lower than PoC-guided cases, demonstrate the framework's capability to synthesize functional exploitation primitives for zero-day-equivalent scenarios where no prior exploit intelligence exists.

The performance differential between cohorts varies by vulnerability class and primitive chain complexity. For canonical OOB vulnerabilities like CVE-2024-53197, the gap is narrower (68\% vs. 85\% for PoC-guided OOB cases) because systematic characterization of out-of-bounds access patterns through PODE S7 offset analysis and S1 KASAN validation provides sufficient grounding even without PoC hints. However, for complex multi-stage chains involving temporal coordination or cross-cache manipulation, the absence of PoC documentation creates more severe degradation: CVE-2024-39486, a Race-to-UAF in IPv6 MLD subsystem, achieves only 45\% SSR compared to 72\% for PoC-guided Race cases, as the framework must infer race window timing through statistical iteration without reference timing primitives.

\end{document}